\documentclass{aa}  
\usepackage{graphicx}
\usepackage{txfonts}
\usepackage{natbib}
\begin{document} 

   \titlerunning{mass loss} 
   \title{New predictions for radiation-driven, steady-state mass-loss and wind-momentum from hot, massive stars}
   \subtitle{I. Method and first results}

   \author{J.O. Sundqvist\inst{1,3}\and
     R. Bj{\"o}rklund\inst{1} \and
     J. Puls\inst{2} \and
     F. Najarro\inst{3} }

   \institute{KU Leuven, Instituut voor Sterrenkunde, Celestijnenlaan 200D, 3001 Leuven, 
   Belgium\ \email{jon.sundqvist@kuleuven.be}\and LMU M\"unchen, Universit\"atssternwarte, Scheinerstr. 1,
     81679 M\"unchen, Germany\and Centro de Astrobiologia, Instituto Nacional de Tecnica Aerospacial, 28850 Torrejon de Ardoz, Madrid, Spain}
                                 
   \date{Received 27/08/2019; accepted 14/10/2019}

 
  \abstract
  {Radiation-driven mass loss plays a key role in the life-cycles of massive stars. However, basic  
  predictions of such mass loss still suffer from significant quantitative uncertainties.}
  {We develop new radiation-driven, steady-state wind models for massive stars with hot surfaces, suitable for quantitative predictions of global parameters like mass-loss and wind-momentum rates.}
  {The simulations presented here are based on a self-consistent, iterative grid-solution to the 
  spherically symmetric, steady-state equation of motion, using full NLTE radiative transfer solutions in the co-moving frame to derive the radiative acceleration. We do not rely 
   on any distribution functions or parametrization for computation of the line force responsible for the wind driving. 
   The models start deep in the subsonic and optically thick atmosphere and extend up to a large radius at which the terminal wind speed has been reached.  
   }
  {In this first paper, we present models representing two prototypical 
  O-stars in the Galaxy, one with a higher stellar mass $M_\ast/M_\odot = 59$ and    
  luminosity $\log_{10} L_\ast/L_\odot = 5.87$ (spectroscopically an early O supergiant)
  and one with a lower $M_\ast/M_\odot = 27$ and 
  $\log_{10}  L_\ast/L_\odot = 5.1$ (a late O dwarf). 
  For these simulations, basic predictions for global 
  mass-loss rates, velocity laws, and wind momentum are given, and the influence from additional parameters like wind clumping and microturbulent speeds discussed. A key result is that although our mass-loss rates agree rather well with alternative models using co-moving frame radiative transfer, they are significantly lower than those predicted by the 
  mass-loss recipes normally included in models of massive-star evolution.}
  {Our results support previous suggestions that Galactic O-star mass-loss rates may be overestimated in present-day stellar evolution models, and that new rates thus might be needed. Indeed, future 
 papers in this series will incorporate our new models into such simulations of stellar evolution, 
 extending the very first simulations presented here toward larger grids covering 
 a range of metallicities, B supergiants 
 across the bistability jump, and possibly also Wolf-Rayet stars.}

   \keywords{radiation: dynamics – radiative transfer – hydrodynamics – stars: massive – 
   stars: mass loss – stars: winds, outflows} 

   \maketitle
%

\section{Introduction}
\label{intro}

For hot, massive stars of spectral types OB, scattering and
absorption in spectral lines transfer momentum from the star's intense
radiation field to the plasma, and so provides the force necessary to
overcome gravity and drive a stellar wind outflow (\citealt{Lucy70}; \citealt{Castor75}; 
review by \citealt{Puls08}). The line-driven winds of OB stars are very strong
and fast, exhibiting mass-loss rates $\dot{M} \sim 10 ^{-5 \dots -9} \, \rm M_\odot/yr$
and terminal wind speeds $\varv_\infty$ that can reach 
several thousand km/s. And since
the radiation acceleration $g_{\rm rad}$ driving these outflows is dominated by
lines from metal ions like Fe, C, N, O, etc., the winds are
predicted, and shown, to depend significantly on stellar metallicity
$Z_\star$ \citep[e.g.,][]{Kudritzki87, Vink01, Mokiem07b}. 

The first quantitative description of 
such line-driven winds was given in the seminal paper 
by \citet{Castor75}, "CAK". By using the so-called Sobolev
approximation\footnote{The Sobolev approximation assumes 
number densities and source functions are constant over a few Sobolev lengths
$L_{\rm Sob} = \varv_{\rm th}/(d\varv_{\rm s}/ds)$ in direction $s$ for ion
thermal speed $\varv_{\rm th}$. This then allows for a \textit{local}
treatment of the radiative line transfer.} in combination with an assumed 
power-law distribution of spectral line-strengths, CAK estimated the total line
force, derived a parameterization for $g_{\rm rad}$, 
and developed a wind-theory that provided very useful 
scaling relations of $\dot{M}$ 
and $\varv_\infty$ with fundamental parameters like stellar 
luminosity $L_\star$ and mass $M_\star$. 

Extensions of this elegant theory 
\citep{Pauldrach86, Friend86} have had considerable
success in explaining many basic features of stellar winds from
OB-stars. However, over the past years evidence has accumulated that such models may not be able to 
make quantitative predictions with the required accuracy. 
Namely, although CAK-type and Sobolev-based Monte-Carlo line-transfer 
models typically agree rather well on predicted $\dot{M}$ \citep{Pauldrach01, Vink00, 
Vink01},  more recent simulations based on line radiation transfer performed 
in the co-moving frame (CMF), or by means of a non-Sobolev Monte-Carlo line 
force, seem to suggest overall lower rates \citep{Lucy07a, Lucy10, Krticka10, Krticka17, Sander17}. 
Moreover, although the studies above, as well as this paper, focus solely on the global, 
steady-state wind, some time-dependent models with 
$g_{\rm rad}$ computed in the observer's frame also indicate similar 
mass loss reductions as compared to CAK-based simulations 
\citep{Owocki99}. 

Concerning attempts to derive $\dot{M}$ directly 
from observations of stellar spectra, a multitude of 
results is scattered throughout the literature. A key 
uncertainty regards the effects of a clumped wind, 
which if neglected in the analysis may lead to rates that differ 
by large factors for the same star, depending on which 
spectral diagnostic is used to estimate $\dot{M}$
\citep{Fullerton06}. When accounting properly for 
such "wind clumping", including the light-leakage effects of 
porosity in physical and velocity space, a few first
multiwavelength studies of Galactic O-stars 
\citep{Sundqvist11, Surlan13, Shenar15} suggest mass-loss 
rates that are lower than or approximately equal to 
those predicted by the \citet{Vink00} standard 
mass-loss recipe.  These results also agree well with X-ray \citep{Cohen14} 
and infra-red \citep{Najarro11} studies, as well as with the 
analysis by \citet{Puls06} that derived upper limits 
on $\dot{M}$ by considering diagnostics ranging from 
the optical to radio. On the other hand, some 
extragalactic studies \citep{Ramirez17, Massa17} find 
upper $\dot{M}$ limits that typically are higher than 
the Vink et al. recipe. Clearly, more effort will 
be needed here in order to place better observational constraints 
on the mass-loss rates from hot, massive stars. 

Reduced mass-loss rates could further have 
quite dramatic consequences for model 
predictions of stellar evolution and feedback
 \citep[e.g.,][]{Smith14, Zsolt17}. Indeed, the 
presence of mass loss has a deciding
impact on the lives and deaths of massive stars, affecting their
luminosities, chemical surface abundances, rotational velocities, and
nuclear burning life-times, as well as ultimately determining which
type of supernova the star explodes as and which type of remnant it leaves behind 
\citep{Meynet94, Langer12, Smith14}. As such, a key aim of this series of papers is -- in addition to 
providing new quantitative predictions of mass loss  
-- to incorporate our new models directly into simulations of massive-star evolution. 

In this first paper, we outline the basic methods of our new wind simulations, 
computed using steady-state hydrodynamics and NLTE\footnote{NLTE=non local thermodynamic equilibrium, which 
here means number densities computed assuming statistical equilibrium 
(see book by \citealt{Hubeny14} for an extensive overview).} CMF radiative transfer calculations 
within the computer-code network {\sc fastwind} 
\citep{Santolaya97, Puls05, Carneiro16, Puls17, Sundqvist18b}. Moreover, we 
present and discuss some first results from two selected models of 
O-stars in the Galaxy. The second paper will present results from 
a larger grid of such O-star simulations, covering metallicities of 
the Galaxy and the Magellanic 
Clouds (Bj{\"o}rklund et al., in prep.). 
The third paper will then extend these calculations to B-stars, examining in 
particular mass-loss and wind-momentum properties over the so-called 
"bi-stability jump" (where iron recombines from being three to two times 
ionized, thus providing much more driving lines, \citealt{Pauldrach90, Vink00}). 
In a fourth paper we will incorporate fully our predictions into models 
of stellar evolution, and carefully investigate the corresponding 
effects (Bj{\"o}rklund et al., in prep.). Further 
follow-up studies will then focus on, e.g., time-dependent effects 
and wind-clumping (see also $\S$\ref{clumping}) 
and potentially even an extension of our current models to 
the highly evolved, classical Wolf-Rayet stars whose 
supersonic winds are optically thick also 
for optical continuum light. 



\section{Basic equations and assumptions} 

Global models (containing both the subsonic stellar photosphere and the supersonic 
wind outflow) are computed by a numerical, iterative 
grid-solution to the steady-state (time-independent) radiation-hydrodynamical 
conservation equations of mass and momentum in spherical symmetry. These are 
supplemented by considering also the energy balance, either by a method based on 
flux-conservation and the thermal balance of electrons \citep{Puls05} or by 
a simplified temperature structure derived from flux-weighted mean 
opacities in a spherical, diluted envelope \citep{Lucy71}.   

For gravitational acceleration $g(r)$ and (isothermal) sound speed 

\begin{equation} 
a^2(r) = \frac{k_{\rm b} T(r)}{\mu(r) m_{\rm h}}, 
\end{equation} 

where 
$T(r)$ is the temperature, $\mu(r)$ the mean molecular weight, $k_{\rm b}$ Bolzmann's constant, and $m_{\rm h}$ the hydrogen atom mass, the equation of motion (e.o.m.) for radial velocity $\varv(r)$ is  
\begin{equation}   
	\varv(r) \frac{d \varv}{dr}(r) \left(1-\frac{a^2(r)}{\varv(r)^2}\right) = g_{\rm rad}(r) - g(r) + \frac{2 a^2(r)}{r} - \frac{da^2}{dr}(r). 
	\label{Eq:eom} 
\end{equation} 
The radiative acceleration 
\begin{equation} 
	g_{\rm rad}(r) = \frac{\kappa_{\rm F}(r) F(r)}{c} = \frac{\kappa_{\rm F}(r) L_\ast}{4 \pi r^2 c}, 
\end{equation} 
where stellar luminosity $L_\ast$ is a fundamental input parameter of the model, $c$ is the speed of light,  
and the flux-weighted mean opacity $\kappa_F$ ($\rm cm^2/g$):  

\begin{equation} 
\kappa_{\rm F} F = \int \kappa_{\nu} F_{\nu} d \nu. 
\end{equation} 

Here opacities $\kappa_\nu$ and radiative fluxes $F_\nu$ are evaluated in 
the frame co-moving with the fluid (the co-moving frame, CMF); to 
order $\varv/c$, this CMF-evaluated $g_{\rm rad}$ can then be used directly in the inertial frame 
e.o.m. eqn. \ref{Eq:eom} (e.g., \citealt{Mihalas78}, their eqns. 15.102-15.113).  The gravity 

\begin{equation} 
g(r) = GM_\ast/r^2, 
\end{equation} 

with gravitation constant $G$, is computed from the second fundamental 
input parameter stellar mass $M_\ast$. The necessary scaling radius for setting 
up the adaptive radial mesh is obtained as described below, giving a "stellar radius" 

\begin{equation} 
R_\ast \equiv r(\tilde{\tau}_F=2/3)
\end{equation} 

for the spherically modified 
flux-weighted optical depth 

\begin{equation} 
	\tilde{\tau}_F(r) = \int \rho(r) \kappa_F(r) \left( \frac{R_\ast}{r} \right)^2 dr.  
\end{equation} 

The stellar "effective temperature" is then here defined as 

\begin{equation} 
	\sigma T_{\rm eff}^4 \equiv \frac{L_\ast}{4 \pi R_\ast^2},  
\end{equation} 

for the Stefan-Boltzmann constant $\sigma$. Note that this differs, for example, 
from the models by \citet{Sander17}, where the stellar 
radius and effective temperature are defined at a Rosseland (rather than flux-weighted) 
optical depth $\tau_{\rm Ross} = 20$. 

Finally, mass-conservation provides the density structure for a 
steady-state mass-loss rate $\dot{M}$, 
\begin{equation} 
	\rho(r) = \frac{\dot{M}} {4 \pi r^2 \varv(r)}. 
\end{equation} 

For a given (also prescribed) chemical composition and metallicity $Z_\ast$, $g_{\rm rad}(r)$ and $a(r)$ are derived from the density, velocity, and temperature structure by using our radiative transfer and 
model atmosphere computer-code package 
{\sc fastwind} \citep{Santolaya97, Puls05, Carneiro16, Puls17, 
Sundqvist18b}; {\sc fastwind} solves for the population number-density rate equations in statistical equilibrium (typically just called NLTE) within the spherically 
symmetric, extended envelope (containing both the subsonic stellar photosphere and the supersonic wind outflow), including the effects from millions of metal spectral lines on the radiation field. 
In the version of {\sc fastwind} used here, chemical abundances are scaled to the solar values 
by \citet{Asplund09} and a standard helium number abundance
$Y_{\rm He}=n_{\rm He}/n_{\rm H} =0.1$, $T(r)$ is derived either as before (from a flux-correction method in the lower atmosphere and the thermal balance of electrons in the outer, \citealt{Puls05}), or by the simplified approach described in the next section. Most importantly, the radiation field -- in 
particular the $g_{\rm rad}(r)$ term responsible for the wind driving -- is now self-consistently computed from full solutions of the radiative transfer equations in the CMF 
for all included chemical elements and contributing lines
\citep{Puls17}; the resulting $g_{\rm rad}$ is then obtained by direct integrations of the 
CMF frequency-dependent fluxes and opacities. 

\section{Numerical calculation methods} 
\label{numerical} 

Each model starts either from the structure obtained in a previous calculation or from 
a structure as computed in previous versions of {\sc fastwind} \citep[see \S2 in][]{Santolaya97}. 
Regarding the latter (standard) option, the key point is that there a quasi-hydrostatic atmosphere below a transition 
velocity $\varv \approx 0.1 a(T_{\rm eff})$ is smoothly combined with an analytic "$\beta$" 
wind velocity  law $\varv(r) = \varv_\infty (1-b R_\ast/r)^{\beta}$, with $b$ 
a constant derived from the transition 
point velocity. This means that the equation of motion is generally not satisfied above 
$\varv \approx 0.1 a$ and that $\dot{M}$, $\varv_\infty$, and $\beta$ are simply provided as 
input by the user. By contrast, the iterative method presented here computes predictive values for $\dot{M}$ 
and $\varv(r)$ for given fundamental parameters $L_\ast$, $M_\ast$, $R_\ast$, and $Z_\ast$.    

The adaptive radial grid consists of 67 discrete points for computing the NLTE number densities and 
the radiative acceleration $g_{\rm rad}$. Whether or not the radial grid is updated between two successive 
iterations depends on the current distribution of grid points in column mass, velocity, and radius. 
In the low-velocity, optically thick parts of the atmosphere, the grid is adjusted according to a 
fixed distribution in column mass; around the sonic point and in the wind parts, the grid is adjusted 
according to the distribution of velocities and radial points. For the solution of 
the e.o.m. eqn. \ref{Eq:eom}, a denser microgrid is used, by means of simple 
logarithmic interpolations from the coarser grid used in the NLTE network.

\subsection{Velocity and density structure} 

For given values of $g_{\rm rad}(r)$ and $a(r)$, the ordinary differential equation (ODE) eqn. \ref{Eq:eom} is solved by a simple Runge-Kutta method to obtain $\varv(r)$ on the discrete radial mesh. Since the radiative acceleration here is an explicit function of only radius, eqn. 
\ref{Eq:eom} is critical at the sonic point $\varv=a$  (see discussion 
in $\S$\ref{topology}); the velocity gradient there is thus derived 
by applying l'Hospital's rule, and the rest of the velocity 
structure then obtained by shooting up and down from the sonic point. The adaptive upper boundary typically lies between $r/R_\ast = 100 - 140$ and the adaptive lower boundary is defined by computing down to
a fixed column mass $m_c^{\rm tot} = -\int_\infty^{r_{\rm min}} \rho dr = 80$.  Adopting such a fixed $m_c^{\rm tot}$ stabilizes the iteration cycle 
somewhat, and is achieved here by using a linear regression to a Kramer-like parameterization for the flux-weighted opacity in the deepest quasi-static layers satisfying both $\varv \la 0.1$ km/s and $\tilde{\tau}_F > 2/3$ (i.e., only in layers below $R_\ast$ and well below the critical sonic point):  

\begin{equation} 
	\kappa_F \approx \kappa_{\rm e} \times (1 + k_a \rho (a^2/10^{12})^{-k_b}),   	
\end{equation}  

where $a^2$ is in cgs-units, $\kappa_{\rm e}$ is the opacity due to Thomson scattering, and $k_a$ and $k_b$ the corresponding 
fit-coefficients. This is essentially the same method as in the standard version of the {\sc fastwind} code \citep{Santolaya97}, except that we here parametrize in sound speed 
$a$ instead of in temperature $T$. Note that although this parametrization indeed provides quite good fits for the deep layers (see Fig. \ref{Fig:force_bal}), it is essential to 
point out that it can only be applied in regions with very low velocity, where the Doppler 
shift effect upon the opacity is negligible\footnote{Moreover, if we were to extend our models to even deeper layers 
with higher temperatures (the lower boundary in the prototypical early O-star model 
presented in the next section lies at $T \approx 10^5$ K), it would also have to be abandoned in order to capture features of, e.g., the so-called "iron-bump" at $T \approx 1.5-2 \times 10^5$ K, where the shift in ionisation of iron-like elements produces a peak also in the quasi-static opacities.}.     
 
For all layers above $\varv \approx 0.1$ km/s, the $g_{\rm rad}(r)$ computed 
from the CMF solution to the transfer equation is used directly, and so does 
not explicitly depend on density, velocity, or the velocity gradient. Thus we must find another way to update the iterative mass-loss rate than, e.g., the singularity and regularity conditions applied 
in CAK-theory. This mass-loss rate update can be done in various ways, and to this end 
we have tried several options. For example, we have preserved the total column 
mass down to some fixed lower boundary radius between successive iterations 
\citep{Pauldrach86, Grafener05} or, in analogy with time-dependent models of line-driven O-star winds, we have tried fixing the density at some lower boundary radius $r_0$ deep in the atmosphere and let the corresponding velocity $\varv_0$ float and iteratively adjust itself to the overlying wind conditions \citep[e.g.,][]{Owocki88, Sundqvist13}. Just like 
\citet{Sander17}, however, we find that (at least among the methods tried thus far) the most stable way to update 
$\dot{M}$ is to instead consider the error in the force balance at the previous critical point $r_{\rm cr}(\varv = a)$ after a new $g_{\rm rad}$ has been computed for a given velocity and sound-speed structure. Defining   

\begin{equation} 
	f_{\rm rc} = 1 - \frac{2 a^2}{r g} + \frac{da^2}{dr}\frac{1}{g},   
\end{equation}  

this factor should equal $\Gamma = g_{\rm rad}/g$ at the sonic point if 
the e.o.m. eqn. \ref{Eq:eom} is fulfilled. In general 
within the iteration cycle though, this will not be the case unless a 
$\delta \Gamma = f_{\rm rc} - \Gamma$ is added to the current estimate 
of $\Gamma$. Since it is well known \citep{Castor75} that for a line-driven wind 
$g_{\rm rad} \propto 1/\dot{M}^\alpha$, with $\alpha$ some power, we may 
thus update our mass-loss rate for iteration $i+1$ according to 
$\dot{M}_{i+1} = \dot{M}_{i} \, (\Gamma/f_{\rm rc})^{1/\alpha}$. Again quite 
analogous to \citet{Sander17}, we find that it is typically easiest for the stability 
of the iteration-cycle to simply assume $\alpha=1$. We stress, however, that this 
(of course) does not mean that really $g_{\rm rad} \propto 1/\dot{M}$, but only 
that this provides a good way of updating the iterative mass-loss rate towards 
convergence.  

\subsection{Temperature structure}      

The radially dependent sound speed is obtained from a calculation of the temperature structure 
(and a radially dependent mean molecular weight). As mentioned above, the temperature update 
can be done self-consistently by means of a flux-conservation+electron 
thermal balance method. In this case, the calculation is done in parallel with computing 
a new estimate of $g_{\rm rad}$, requiring that both temperature and radiative 
acceleration be converged before the next update of velocity and density. However, 
since converging temperature and radiative acceleration in parallel is very 
time-consuming, and sometimes leads to somewhat un-stable iteration cycles, 
most models here instead update the temperature structure using a simplified 
method. Namely, following \citet{Lucy71} 
we may estimate the temperature structure in the spherically extended 
envelope as 

\begin{equation} 
	T(r) = T_{\rm eff} \left( W(r) + 3 \tilde{\tau}_F(r)/4 \right)^{1/4}
	\label{Eq:Tsimple} 
\end{equation} 
 
for dilution factor $W(r) = 1/2 \left(1- \sqrt{1 - R_\ast^2/r^2} \right)$ and where for radii 
$r < R_\ast$ simply $W = 1/2$. Analogous to the standard version of 
{\sc fastwind}, we also here impose a floor-temperature of typically $T \approx 
0.4 T_{\rm eff}$ in order to prevent excessive cooling in the outermost 
wind (see also \citealt{Puls05}). Given current estimates for $g_{\rm rad}(r)$ and $\mu(r)$, eqn. \ref{Eq:Tsimple} can now be formulated as an ODE for $da^2/dr$ and solved in parallel with the 
e.o.m. eqn. \ref{Eq:eom}. In contrast to the flux conservation+electron thermal 
balance method, this temperature structure 
is then held fixed (along with the new $\varv(r)$ and $\rho(r)$) within the next NLTE iteration loop for obtaining the updated $g_{\rm rad}$. Although this procedure means we do not 
require perfect radiative equilibrium, we have found that the corresponding 
impact on the wind dynamics is small (see also \citealt{Pauldrach86}). As such, the models in 
\S \ref{results} all assume this structure, except for in \S \ref{tlucy} where a detailed comparison 
to a model computed from full flux-conservation+electron thermal 
balance is presented. 

\subsection{Radiative acceleration} 
\label{rad_acc} 

As mentioned in the previous section, the radiative acceleration is derived from the NLTE rate equations 
and full solutions of the radiative transfer equation in the CMF, for all elements from H to Zn and including "all" contributing lines (i.e. all those included in the corresponding model atoms). This is different from previous versions of {\sc fastwind}, aimed at quantitive spectroscopy of optical and infra-red spectral lines, where opacities from elements not included in the detailed spectroscopic investigations (the "background" elements) were added up to form the quasi-continuum quantities used to compute the radiation field. Although the CMF transfer solver thus is a new addition to the code \citep{Puls17}, the corresponding model atoms are the same as in previous versions (coming primarily from the Munich database compiled by \citealt{Pauldrach01}). 

Numerous tests have been performed to verify the validity of the new CMF solver for {\sc fastwind}; for example extensive comparison of "$\beta$-law" models with the well-established {\sc cmfgen} code \citep{Hillier98} shows excellent agreement between the radiative accelerations computed by the two independent programs (Puls et al., in prep., see also Fig. 1 in \citealt{Puls17}).  

The radiative acceleration $g_{\rm rad}$ depends on opacities and fluxes, where the
opacities are computed from a consistent NLTE solution that itself depends 
on the (CMF-)radiation field. After a
consistent solution for the frequency-dependent opacities and fluxes have been obtained, 
the final $g_{\rm rad}$ is derived from direct numerical integrations. 
If the simplified temperature structure (see above) is applied, the procedure is performed for a given velocity, density, and temperature, and the NLTE cycle iterated until $g_{\rm rad}$ is converged. Then this 
$g_{\rm rad}$ is used to compute a new temperature, velocity, and density, and so forth. On the 
other hand, if the flux conservation+electron thermal balance method is applied, the temperature
is computed in parallel with $g_{\rm rad}$, and the next hydrodynamic update of 
density and velocity done only when both have converged. Particularly when quite big changes between successive hydrodynamic iterations are present, this often leads to more unstable iteration-cycles and 
significantly slower model convergence.   

Finally, just like in previous versions of {\sc fastwind} 
(and as in basically all other 1D model atmosphere codes), we broaden all spectral line-profiles with 
an additional isotropic "microturbulent" velocity $\varv_{\rm turb}$, set here to a standard  
O-star value 10 km/s (but see discussion in \S \ref{vturb} regarding the impact of this parameter on the wind dynamics). Note that this $\varv_{\rm turb}$ also enters the calculations of the NLTE 
number densities and opacities, due to its influence on the radiative interaction rates. 

\subsection{Convergence criteria}  

The basic convergence criterion for our models is that the e.o.m. be fulfilled after a 
new $g_{\rm rad}(r)$ has been calculated from a given density and velocity structure. Re-writing 
the e.o.m. eqn.~\ref{Eq:eom} as 

\begin{equation} 
	1 - \frac{\varv \frac{d \varv}{dr} \left(1-\frac{a^2}{\varv^2}\right) - g + \frac{2 a^2}{r}	  
	- \frac{da^2}{dr}}{g_{\rm rad}} = f_{\rm err}  
\label{Eq:ferr} 
\end{equation} 
we note that $f_{\rm err}$ in eqn. \ref{Eq:ferr} is zero in a dynamically consistent model. We thus require that  

\begin{equation} 
	f_{\rm err}^{\rm max} = \rm max( \, abs(f_{\rm err}) \,)  
	\label{Eq:max_err} 
\end{equation} 

be below a certain threshold, typically set to $0.5 -1 \times 10^{-2}$ (and 
neglecting the deepest quasi-static layers where the Kramer-like approximation is used for $g_{\rm rad}$, see 
above); this means that for a converged model the e.o.m. is everywhere fulfilled to 
within better than a percent. 

In addition, we also make sure that sufficient 
convergence (again typically better than a percent) is obtained for the
hydrodynamic variables $\dot{M}$, $\varv$ and $T$, using 
the maximum relative change between two successive iterations $i-1$ and $i$. 
That is, for quantity X we require that  

\begin{equation} 
\rm  \Delta X_i = max( \, abs( \, X_i/X_{i-1}-1 \,) \,) 
\label{Eq:delta_x} 
\end{equation}

be below $\approx 0.01$. 

A key point with the above is that the main convergence criterium 
eqn. \ref{Eq:max_err} is applied directly on the actual e.o.m. Since the 
e.o.m. then per definition is fulfilled for a converged model, this 
means that we should avoid some potential difficulties related to "false 
convergence", which could (at least in principle) occur for criteria 
involving only relative changes between iterations of the hydrodynamic 
variables (such as eqn. \ref{Eq:delta_x}). On the other hand, there is nothing in the above 
that guarantees that the converged solution is also unique; this  
is further discussed in $\S$\ref{topology}.

\section{First results}
\label{results}

\begin{table*}
\begin{minipage}{\textwidth}
    \centering
    \caption{Input stellar and predicted wind parameters of models.}
        \begin{tabular}{ l l l l l l l l l}
        \hline \hline 
        Model    & $M_\ast/M_\odot$ &  $\log L_\ast/L_\odot$  & $R_\ast/R_\odot$ 
                     & $T_{\rm eff}$ [K] & $Z/Z_\odot$ & $\log \dot{M}/\rm M_\odot/yr$ & $\varv_\infty$ [km/s]  & 
                     $\eta = \dot{M} \varv_\infty c/L_\ast$ \\
                     \hline
        "early" O  & 59.3 & 5.87 & 18.0  & 40\,000 & 1.0 & -5.82 & 2\,480 & 0.25 \\
        "late" O  & 26.6 & 5.10 & 9.37  & 35\,530 & 1.0 & -7.59 & 5\,300 & 0.05 \\
                \hline
        \end{tabular}
    \label{Tab:params}
\end{minipage}
\end{table*}

\begin{figure}
\begin{minipage}{8.5cm}
\resizebox{\hsize}{!}
            {\includegraphics[angle=90]{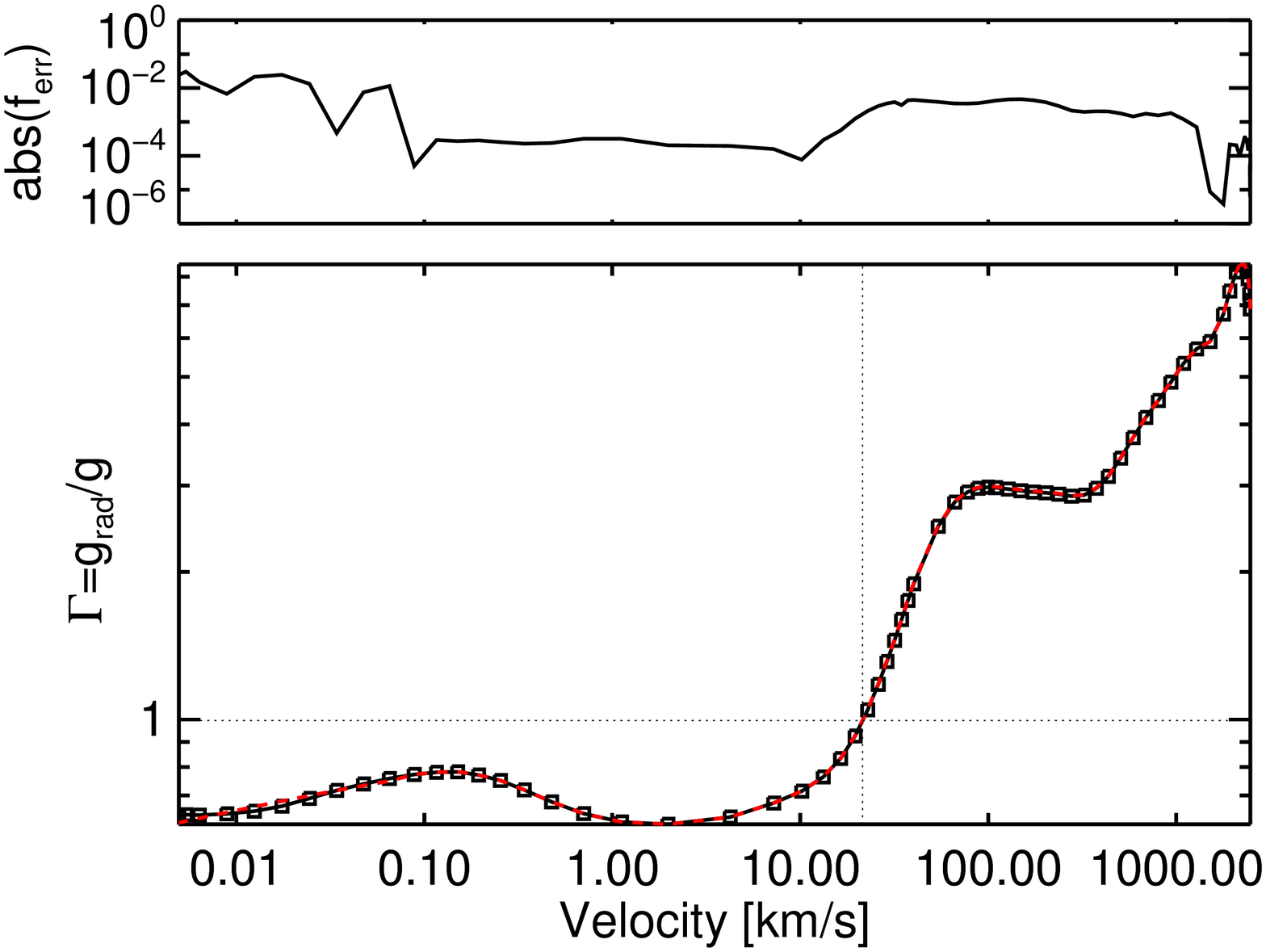}}
\end{minipage} 
\begin{minipage}{8.5cm} 
\resizebox{\hsize}{!}
            {\includegraphics[angle=90]{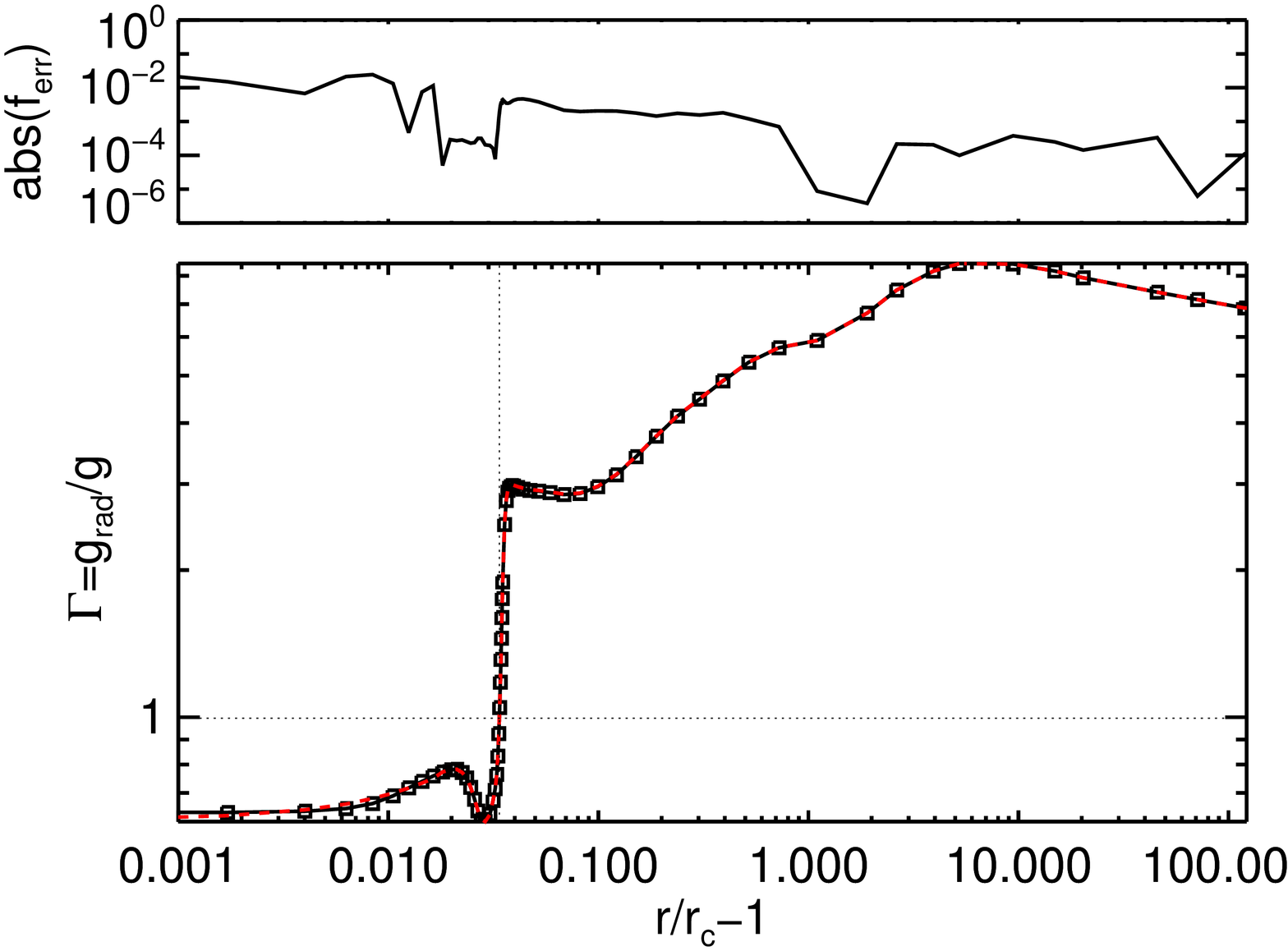}}
  \end{minipage} 
  \caption{Final force balance for the early O-star model in Table 1. In the lower panels 
  of both figures, the solid lines and black
  squares show $\Gamma = g_{\rm rad}/g$ on the discrete mesh points and the red dashed line values 
  for the rest of the terms in the e.o.m., eqn. \ref{Eq:eom}, i.e., $(1 -2a^2/r 
  + da^2/dr + \varv d \varv/dr(1-a^2/\varv^2))/g$. Since the e.o.m. is perfectly 
  fulfilled, the solid black lines and the dashed red ones lie directly on top of each other in the figure.
  The dotted lines mark the sonic point 
  $\varv = a$. The upper panels then display the absolute value of the error $f_{\rm err}$ (eqn. \ref{Eq:ferr}) in the e.o.m. The abscissae in the upper 
  figure show velocity $\varv$ and in the lower scaled radius 
  $r/r_c -1$, with $r_c$ the lower boundary radius of the simulation.}
  \label{Fig:force_bal}
  \label{Fig:force_bal_rad}
\end{figure}

\begin{figure}
\begin{minipage}{8.5cm} 
\resizebox{\hsize}{!}
            {\includegraphics[angle=90]{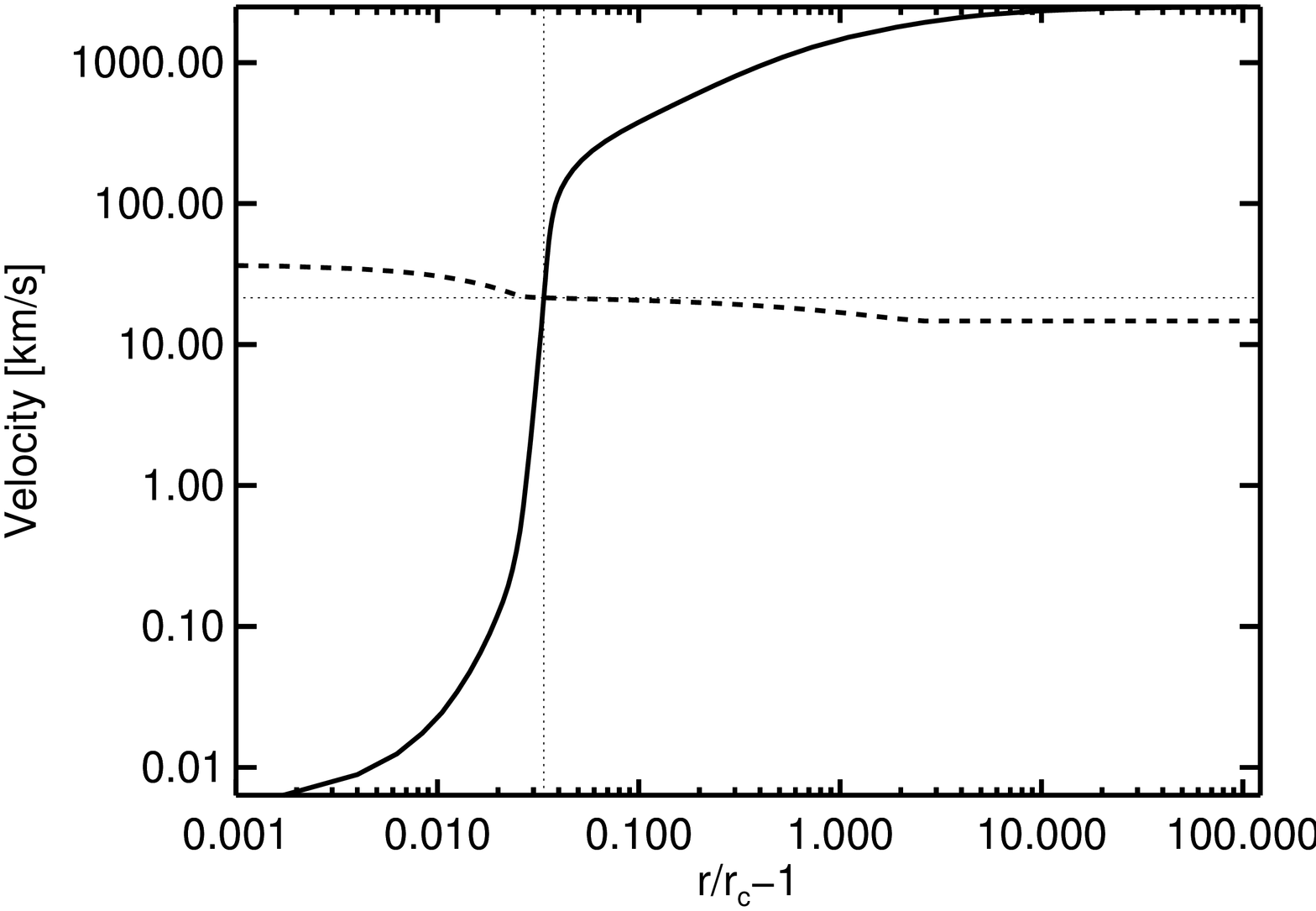}}
\end{minipage} 
\begin{minipage}{8.5cm} 
\resizebox{\hsize}{!}
    {\includegraphics[angle=90]{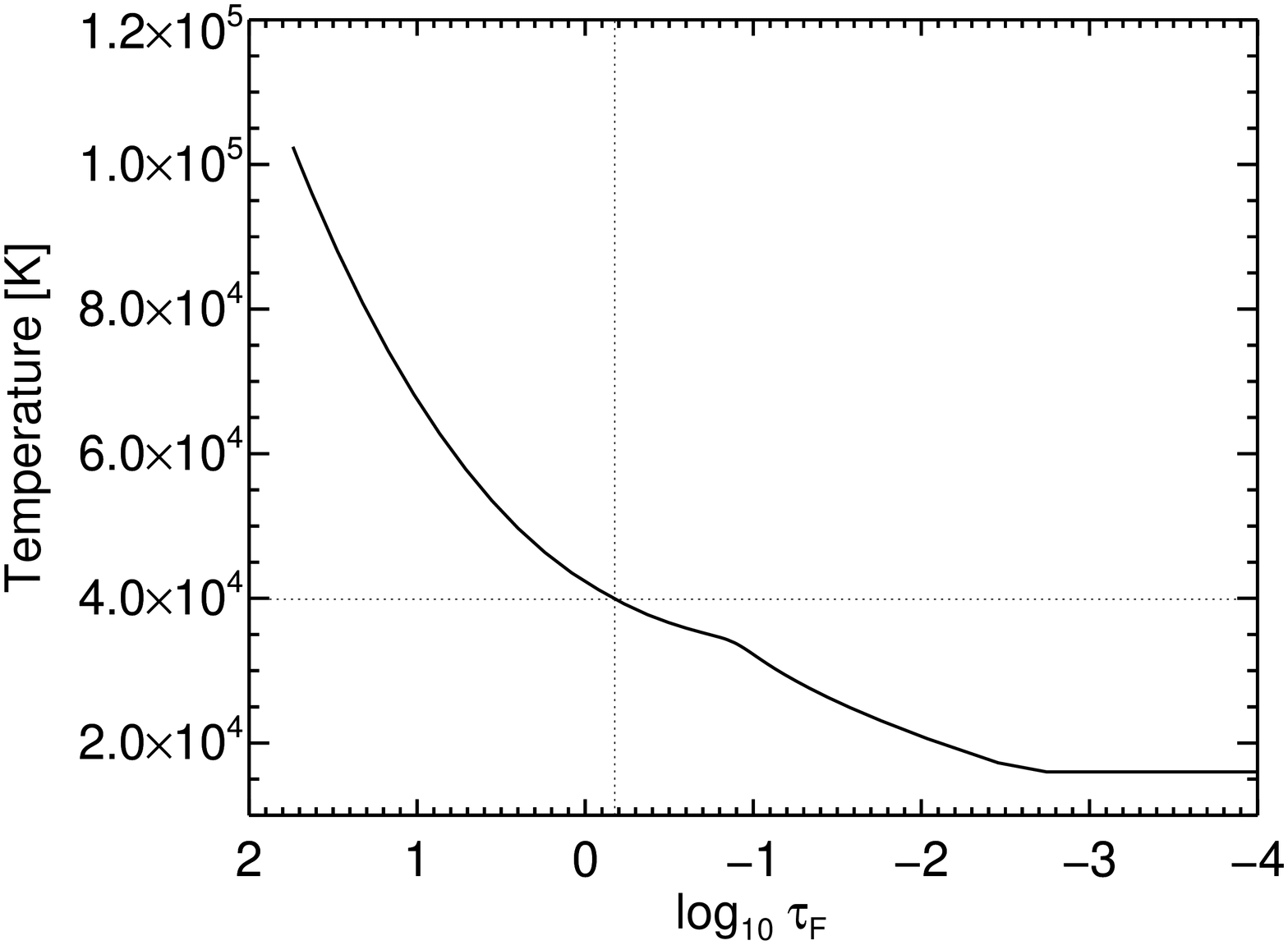}}
\end{minipage} 
  \caption{ \textbf{Upper panel:} Velocity $\varv$ (solid line) and sound speed $a$ (dashed line) as function of scaled radius $r/r_c -1$,  
  with $r_c$ the radius at the lower boundary, for the early O-star model in Table 1. The dotted lines mark the position at which $\varv=a$. \textbf{Lower panel:} Temperature as function of spherically modified flux-weighted optical depth $\tilde{\tau}_F$. The dotted lines mark the position at which $\tilde{\tau}_F=2/3$ (and thus $T=T_{\rm eff}$, see text).}
  \label{Fig:v_law}
    \label{Fig:t_law}
\end{figure}  
 
\begin{figure*}
\begin{center}
\begin{minipage}{12.5cm}
\resizebox{\hsize}{!}
            {\includegraphics[angle=90]{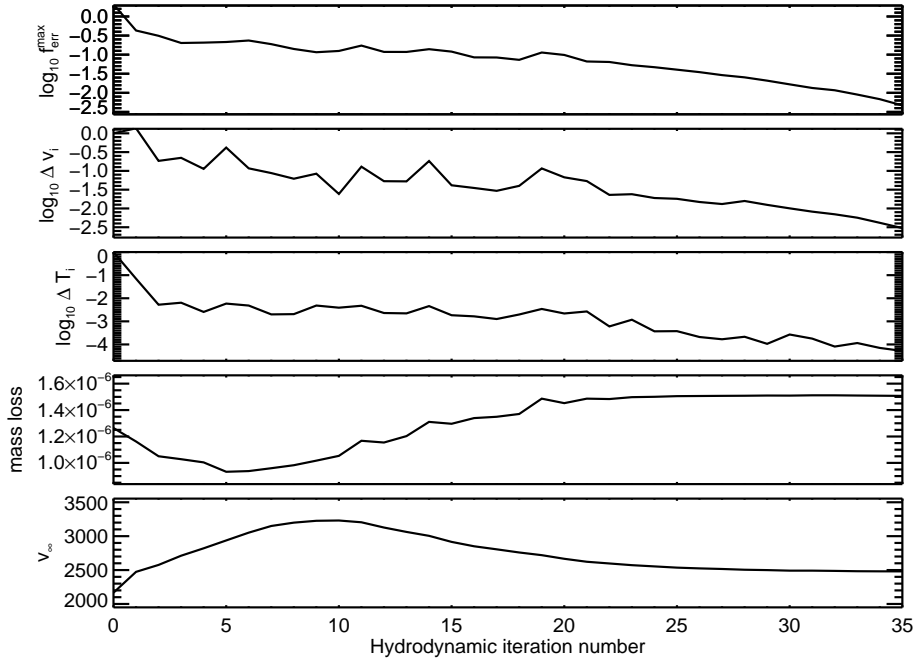}}
  \caption{Iterative evolution of quantities vs. hydrodynamic iteration number $i$. The uppermost panel shows the maximum error in the e.o.m.; the second panel the relative change in velocity; the third the relative change in temperature; the fourth the mass-loss rate; and the fifth the terminal wind speed. See text.}
  \label{Fig:err_plot}
  \end{minipage} 
    \end{center}
\end{figure*} 
    
\begin{figure}
\begin{minipage}{8.5cm}
\resizebox{\hsize}{!}
            {\includegraphics[angle=90]{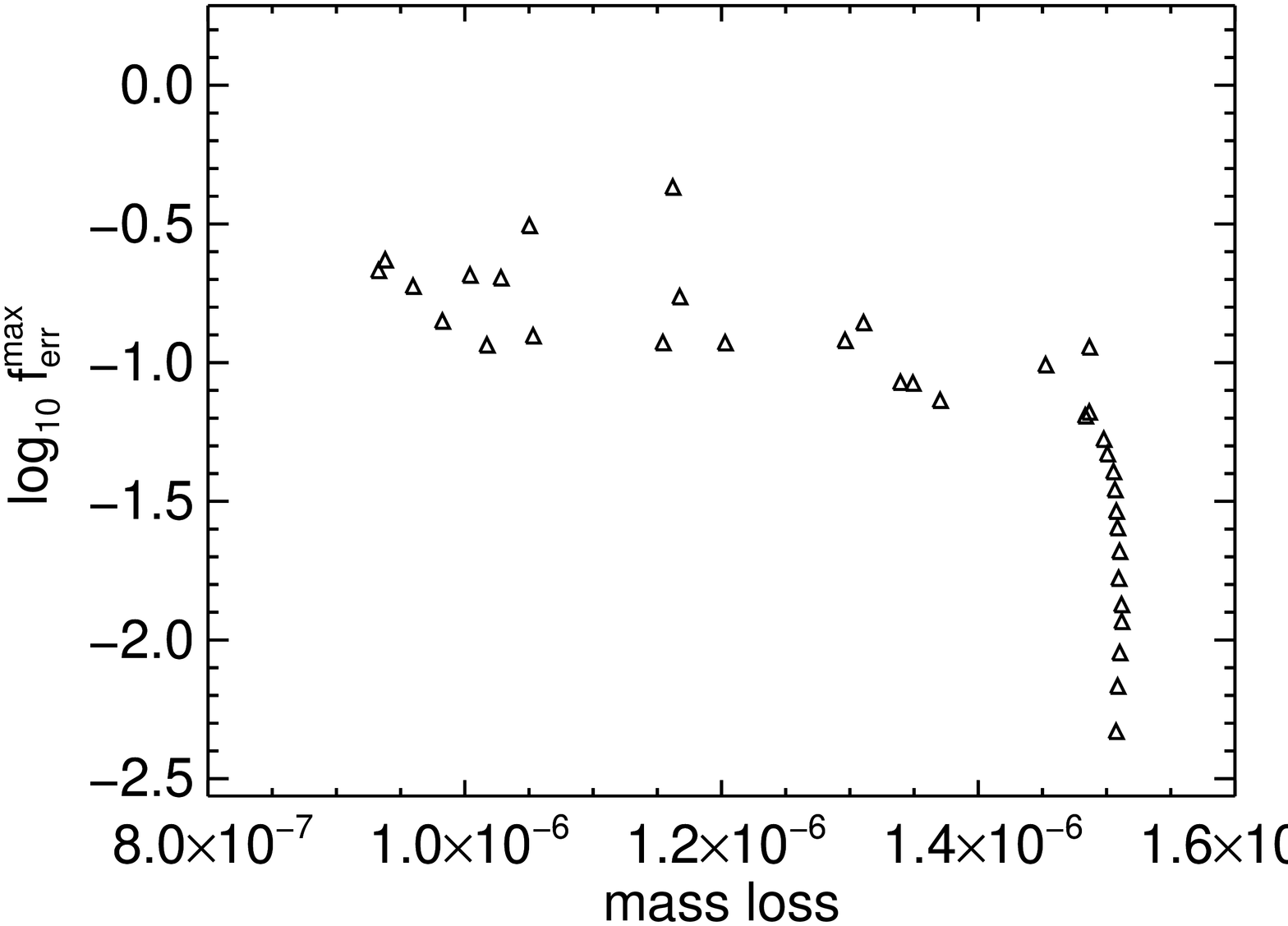}}
\end{minipage}
\begin{minipage}{8.5cm} 
\resizebox{\hsize}{!}
            {\includegraphics[angle=90]{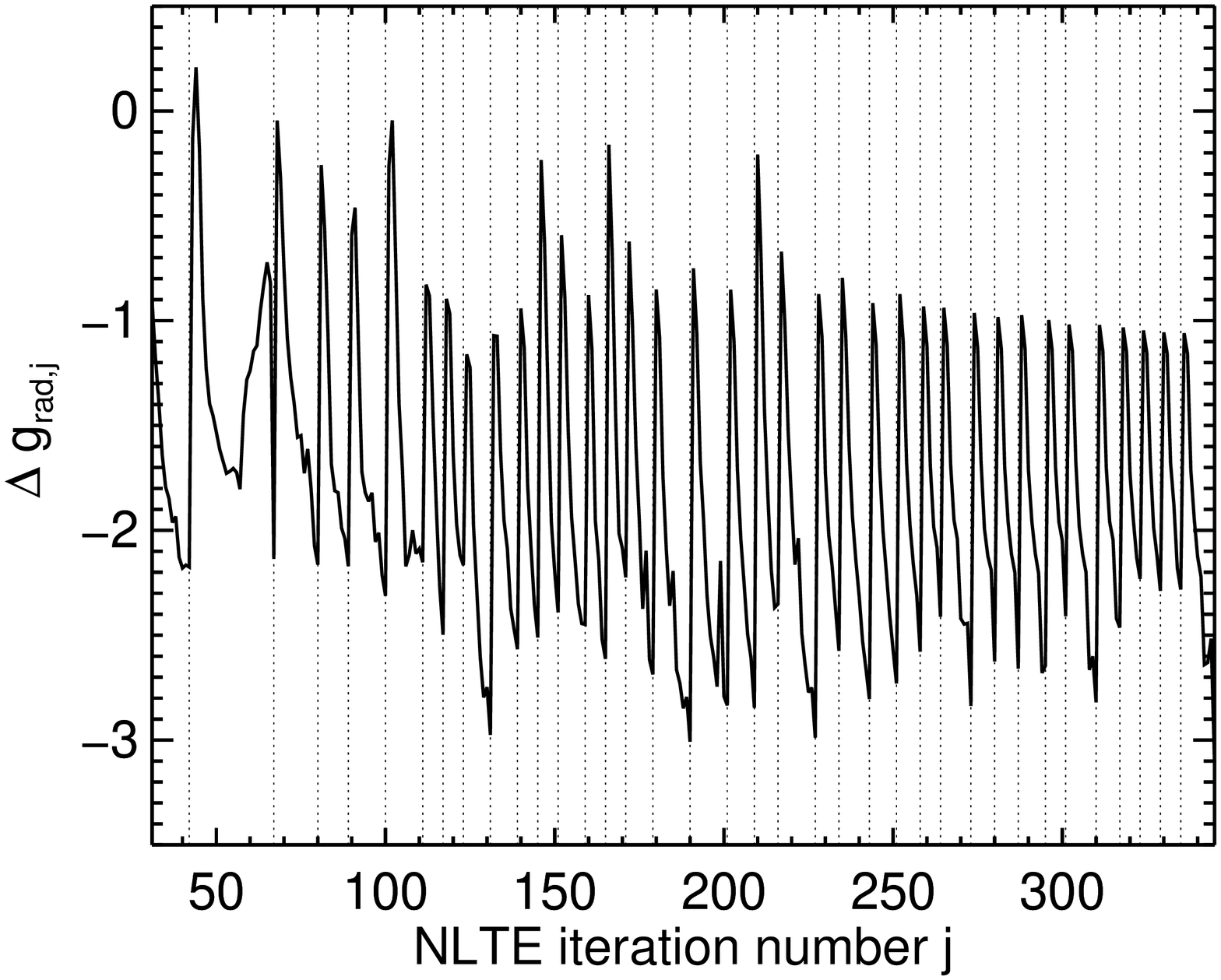}}
\end{minipage} 
  \caption{\textbf{Upper panel:} The triangles show maximum error in the e.o.m. 
  and the corresponding mass-loss rate for each hydrodynamic iteration $i$. 
  \textbf{Lower panel:} The relative change in radiative acceleration between 
  two successive NLTE iterations $j-1$ and $j$. The vertical dotted lines mark 
  NLTE iterations $j$ where hydrodynamic iteration updates $i$ are made. 
  Both panels show results from 
  the early O-star model in Table 1.}
  \label{Fig:ferr}
    \label{Fig:nlte_it}
\end{figure} 

\begin{figure}
\resizebox{\hsize}{!}
            {\includegraphics[angle=90]{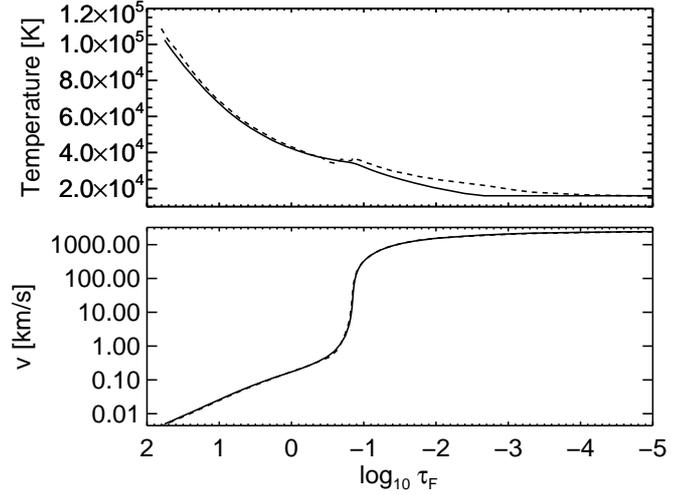}}
  \caption{Comparison of converged temperature (upper panel) and velocity (lower panel) for early O-star models computed with a simplified temperature structure (solid 
  lines) and a flux-conservation+electron thermal balance method (dashed lines).   
  On the abscissae are radially modified flux-weighted optical depth.}
  \label{Fig:t_comp}
\end{figure} 

\begin{figure}
\begin{minipage}{8.5cm}
\resizebox{\hsize}{!}
            {\includegraphics[angle=90]{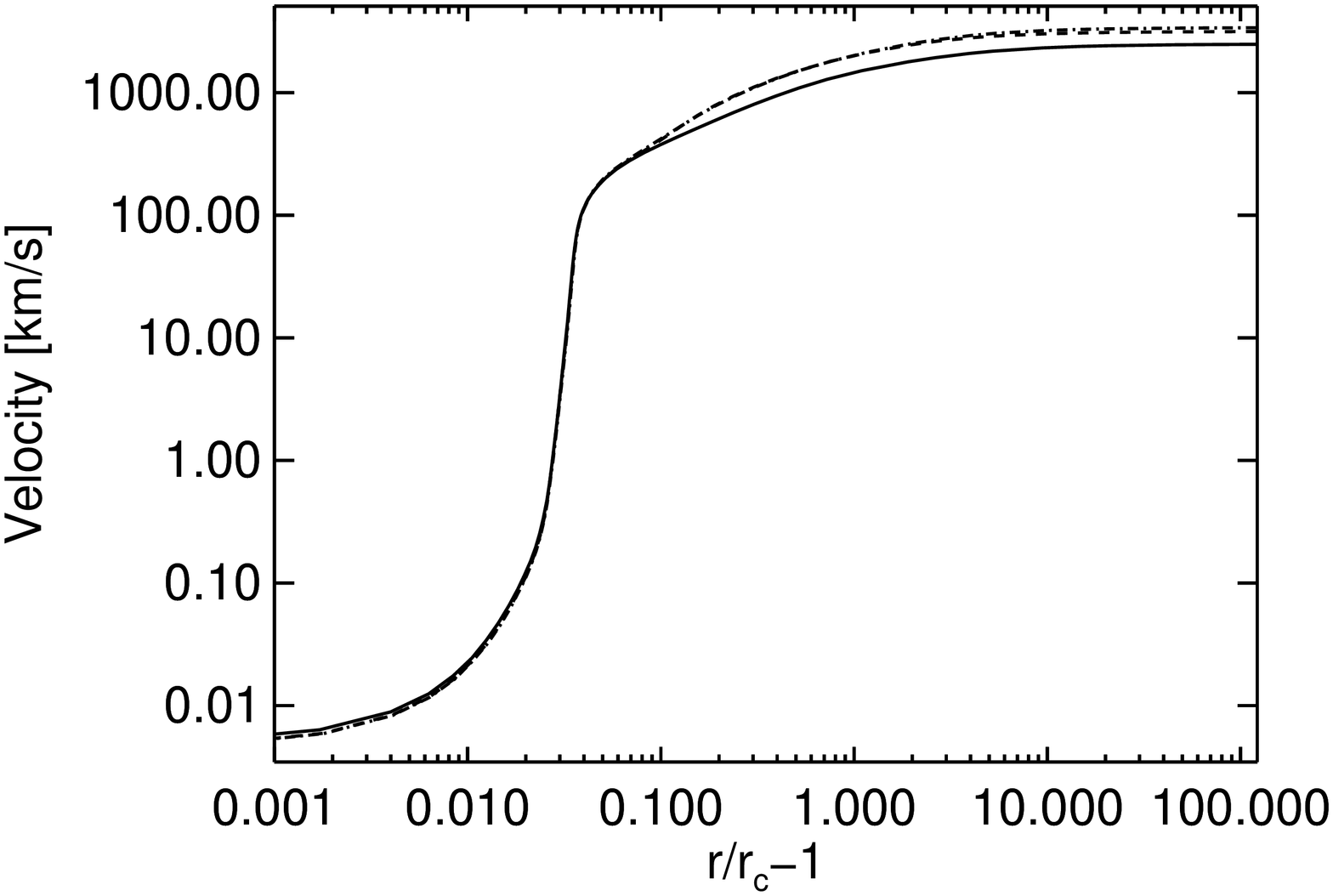}}
\end{minipage}
\begin{minipage}{8.5cm} 
\resizebox{\hsize}{!}
            {\includegraphics[angle=90]{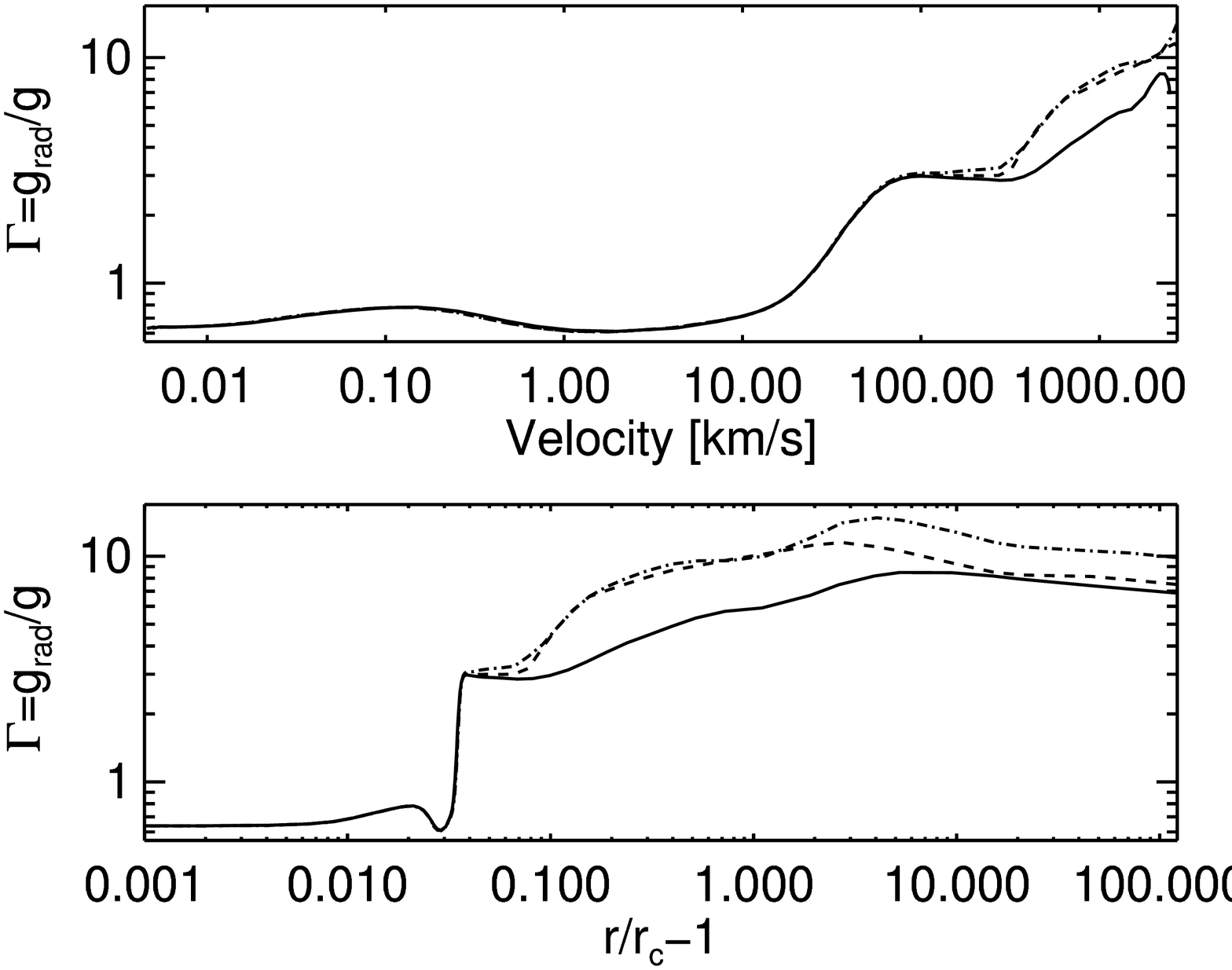}}
\end{minipage} 
  \caption{Comparison of converged structures for early O-star models with no wind clumping 
  and no X-rays (see text) (sold lines), with wind clumping but without X-rays (dashed lines), and 
  with both wind clumping and X-rays (dashed-dotted lines). The uppermost panel shows velocity vs. 
  scaled radius and the lower two panels $\Gamma$ vs. velocity (upper) and scaled    
  radius (lower). A clumping factor $f_{\rm cl} =10$, starting 
  at $\varv = 0.1 \varv_\infty$, is assumed. See text.}
  \label{Fig:clump}
\end{figure} 

\begin{figure}
\resizebox{\hsize}{!}
            {\includegraphics[angle=90]{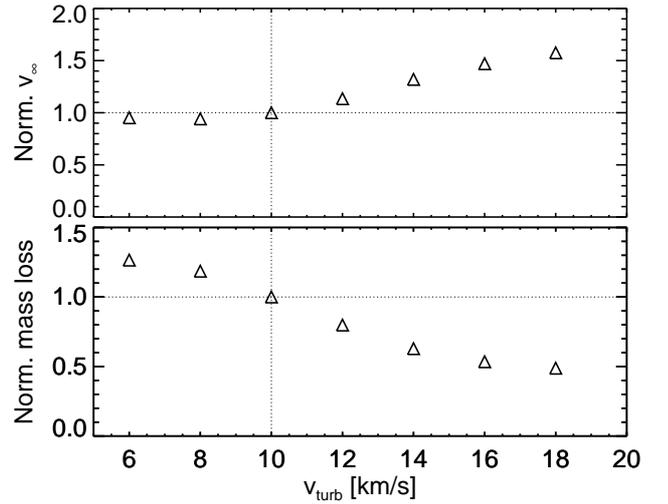}}
  \caption{Comparison of early O-star models computed for different values 
  of "micro-turbulent" velocity $\varv_{\rm turb}$. The upper panel compares the 
  computed terminal wind speeds $\varv_\infty$ 
  and the lower panel the mass-loss rates $\dot{M}$. Both panels 
  have been normalized to the results of the standard model with $\varv_{\rm turb} = 10 \, \rm km/s$.}
  \label{Fig:vturb}
\end{figure} 

\begin{figure}
\begin{minipage}{8.5cm}
\resizebox{\hsize}{!}
            {\includegraphics[angle=90]{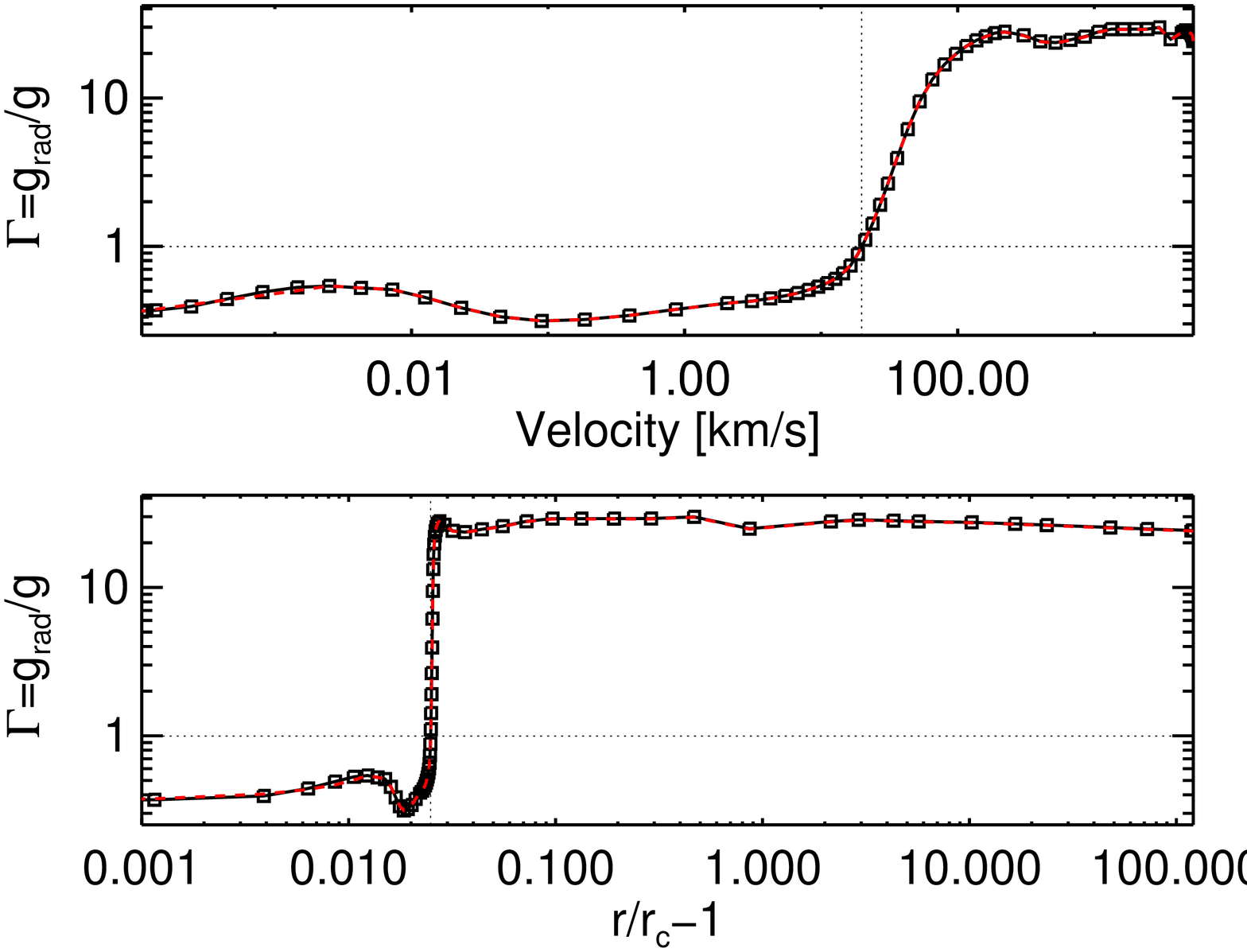}}
\end{minipage}
\begin{minipage}{8.5cm} 
\resizebox{\hsize}{!}
            {\includegraphics[angle=90]{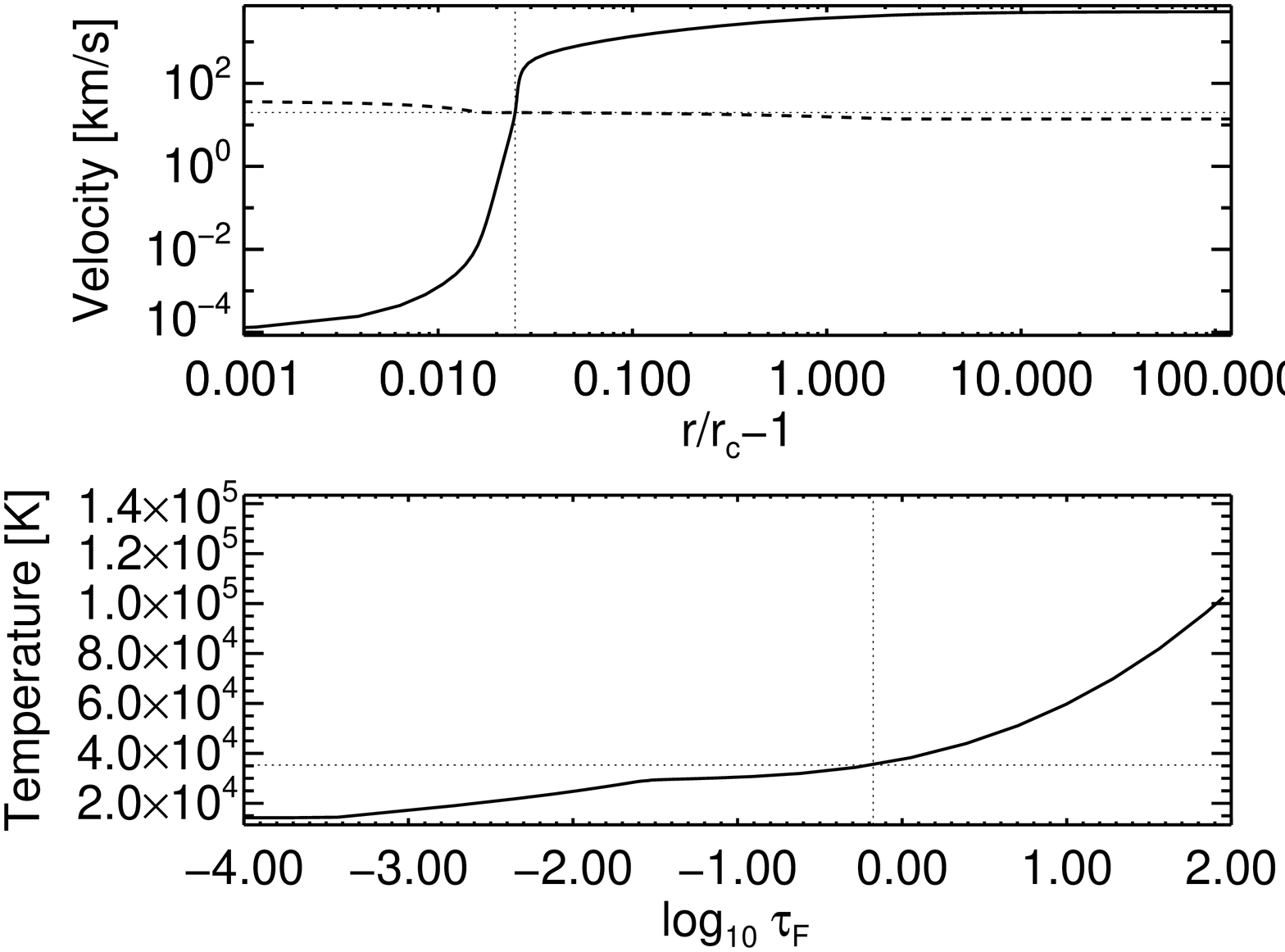}}
\end{minipage} 
  \caption{Final results for the "late" O-star model, with parameters as in Table 1. The 
  upper two panels display the $\Gamma$-factors and the lower two panels 
  the velocity and temperature structures, analogous, respectively, to Figs. \ref{Fig:force_bal} and 
  \ref{Fig:v_law} for the "early" O-star model.} 
  \label{Fig:O7}
\end{figure} 

\begin{figure}
\resizebox{\hsize}{!}
            {\includegraphics[angle=90]{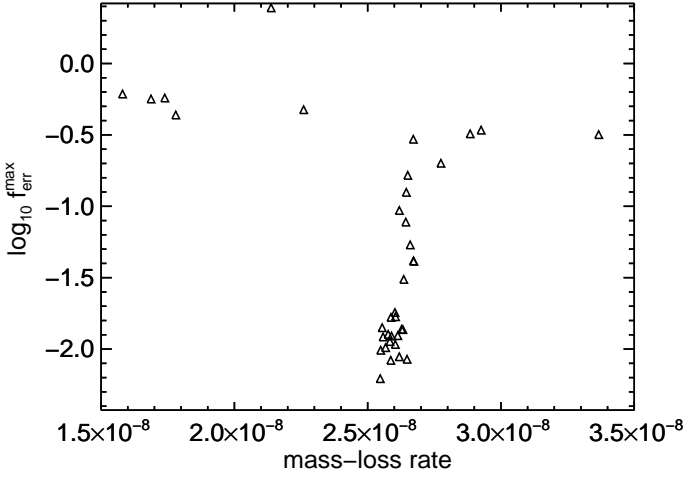}}
  \caption{The triangles show maximum error in the e.o.m. 
  and the corresponding mass-loss rate for each hydrodynamic iteration $i$ 
  for the "late" O-star in Table 1.}
  \label{Fig:O7_ferr}
\end{figure} 

\begin{figure}
\begin{minipage}{8.5cm}
\resizebox{\hsize}{!}
             {\includegraphics[angle=90]{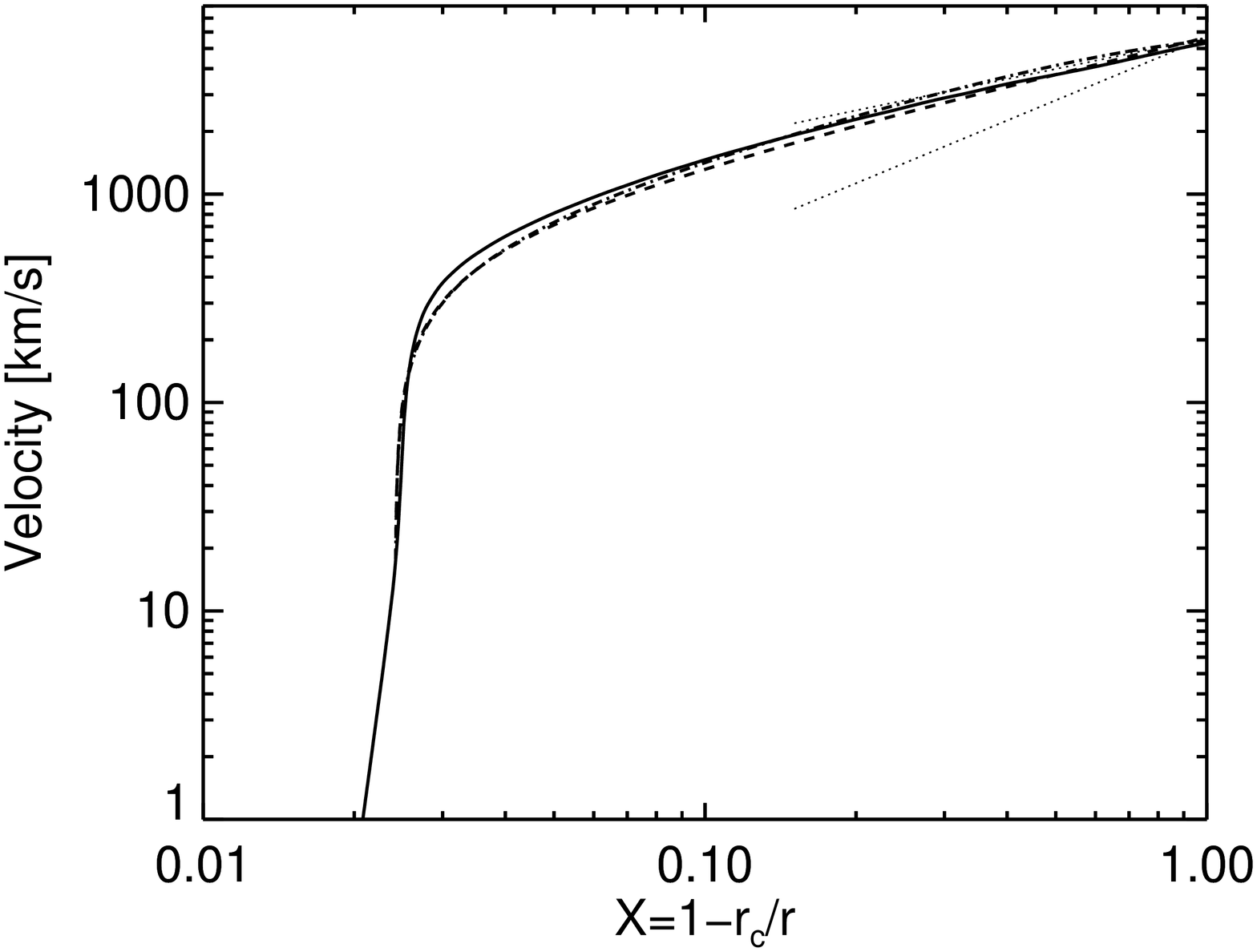}}
\end{minipage}
\begin{minipage}{8.5cm} 
\resizebox{\hsize}{!}
                {\includegraphics[angle=90]{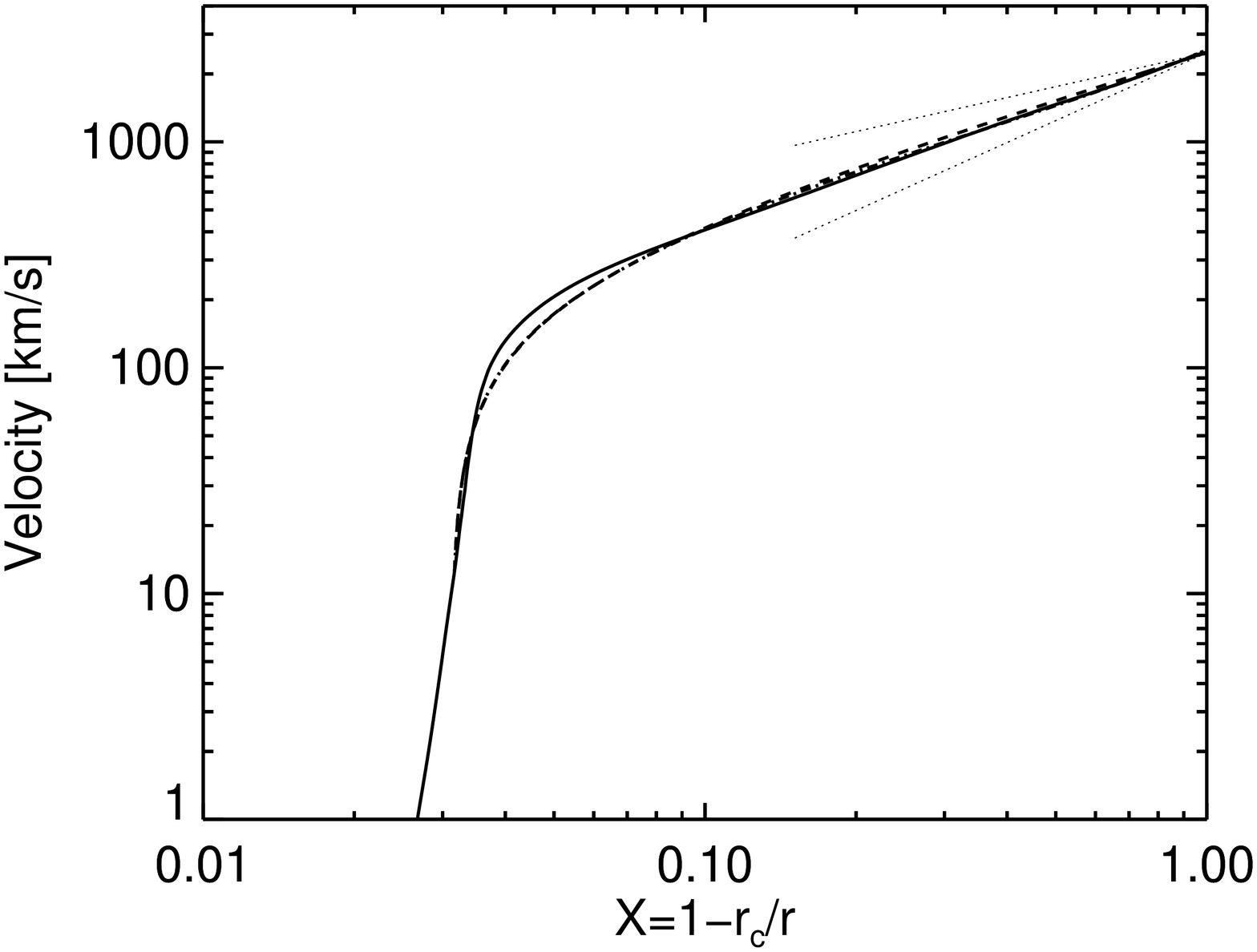}}
\end{minipage} 
  \caption{Comparison of the velocity fields in the 
  self-consistent hydrodynamic simulations (solid lines) to 
  fits assuming "single" (dashed lines) and "double" 
  (dashed-dotted lines) $\beta$-laws. See text.   	
  The dotted background lines compare the fits to simple 
  $\beta=0.5$ and $\beta=1.0$ laws. The upper and lower 
  panels show the late and early O-star models from Table 1, 
  respectively.}
  \label{Fig:beta}
\end{figure} 

\begin{figure}
\resizebox{\hsize}{!}
            {\includegraphics[angle=90]{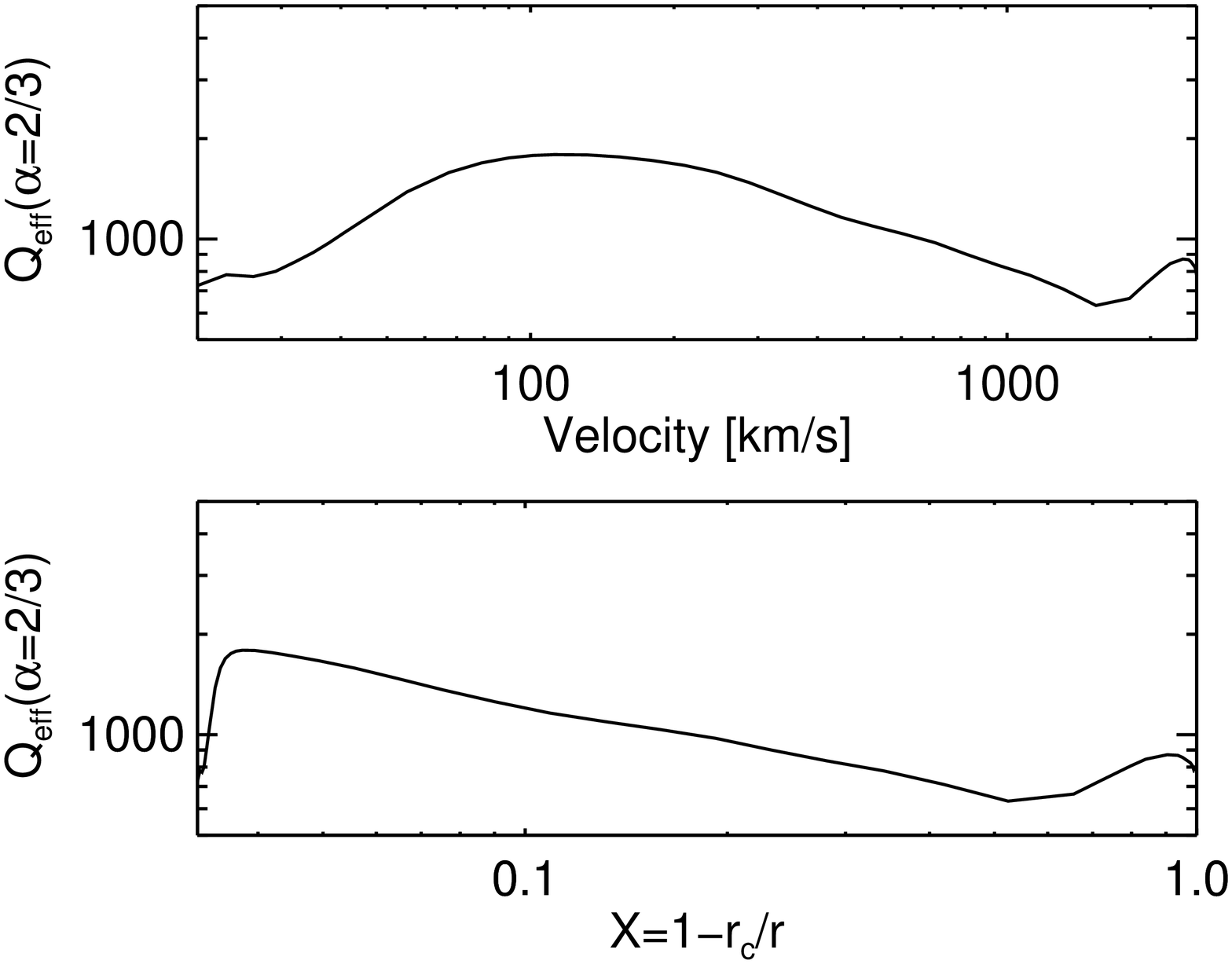}}
  \caption{Effective $Q_{\rm eff}$ values (eqn. \ref{Eq:qeff}) for 
  the converged early O-star model (Table 1) and $\alpha=2/3$, 
  as function of velocity (upper panel) and dimensionless radius parameter 
  $X$ (lower panel).} 
  \label{Fig:qeff}
\end{figure} 

Table 1 displays fundamental stellar parameters and the final predicted 
values for $\dot{M}$ and terminal wind speed $\varv_\infty$ 
(defined here simply as the velocity at the outermost grid-point) 
for two models, representing a prototypical higher-mass 
and lower-mass O-star in the Galaxy; following the 
calibration by \citet{Martins05} the two models would be 
classified as an O4I and O7V star, respectively.     
This section presents results from these two 
base models, starting with a detailed analysis of the former. 

\subsection{Early O-star model} 

For the more luminous, higher-mass model, $M \approx 60 M_\odot$, Fig. 
\ref{Fig:force_bal} displays
the final, converged force balance of the simulation, demonstrating a  
perfect agreement between left and right hand sides in the e.o.m eqn. 2. 
We note further that, actually, even the lowermost part where the 
Kramer-like approximation (see above) is applied gives a force balance 
accurate to within a few percent (see upper panel) for the final 
fit-parameters $k_a = 10^{13.19}$ and $k_b = 4.611$.  

The predicted mass-loss rate of the model is $1.51 \times 10^{-6} \rm \, M_\odot/yr$, 
the terminal wind speed is 2480 km/s, and the scaled wind-momentum rate 
$\eta = \dot{M} \varv_\infty c/L_\ast = 0.25$. Above the quasi-static deep layers 
the velocity field (illustrated in Fig. \ref{Fig:v_law}) displays a very 
steep acceleration around the sonic point, followed 
by a region of slower acceleration around $\sim100$ km/s. 
Due to the (here implicit) 
dependence of $g_{\rm rad}$ on the velocity gradient in these 
highly supersonic regions, this feature also causes the corresponding radiative acceleration 
plateau visible in Fig. \ref{Fig:force_bal} around $\sim 100$ km/s. 
The 
radially modified flux-weighted optical depth at the sonic point 
is $\tilde{\tau}_F \approx 0.14$ and the stellar radius (per definition 
at $\tilde{\tau}_F =2/3$) is located at a velocity $\varv \approx 0.2$ km/s, 
i.e. well below the sonic point. We further also note the dip in radiative acceleration in 
the atmospheric regions leading up to the sonic point (see 
Fig. \ref{Fig:force_bal_rad}). This is qualitatively similar to the 
behavior observed in some earlier non-Sobolev simulations
based on simplified methods to compute $g_{\rm rad}$  (\citealt{Owocki99}), 
however further studies are required to determine  more quantitatively 
potential connections to those models. Fig. \ref{Fig:t_law} displays the 
accompanying temperature structure, confirming that 
$T(\tilde{\tau}_F=2/3) = T_{\rm eff}$ but also illustrating that, of 
course, no temperature reversal (see \S \ref{tlucy}) is seen in this model computed 
by means of the simplified eqn. \ref{Eq:Tsimple}. 

Fig. \ref{Fig:err_plot} displays some characteristics of the iteration cycle towards 
hydrodynamic convergence. The top panel shows the maximum 
error in the e.o.m., $f_{\rm err}^{\rm max}$ (eqn. \ref{Eq:max_err}), the second 
and third panels the maximum relative change in 
velocity and temperature between two iterations (eqn. \ref{Eq:delta_x}), 
and the two bottoms panels give the iterative evolution of mass-loss rate 
and terminal wind speed. The figure demonstrates
how, as the error in the e.o.m. decreases, the mass-loss rate, velocity 
field, and temperature structure all stabilize toward convergence. 
Fig. \ref{Fig:ferr} displays the combination of $f_{\rm err}^{\rm max}$ 
and $\dot{M}$ for each iteration $i$, showing again how after an 
initial adjustment period the mass-loss rate remains fairly constant 
while $f_{\rm err}^{\rm max}$ decreases towards convergence.    
Note again that for each such hydrodynamic 
iteration-update $i$ the NLTE and CMF radiative transfer 
computations iterate toward a new converged radiative acceleration for 
the given $\varv(r)_i$, $\rho(r)_i$, and $T(r)_i$. Fig. \ref{Fig:nlte_it} 
demonstrates this, showing $\Delta g_{\rm rad,j}$ over the NLTE iteration 
cycle $j$; in this figure, the valleys in $\Delta g_{\rm rad,j}$ correspond 
to the places at which $g_{\rm rad}$ has relaxed and 
a new update $i$ of the hydrodynamic variables is performed.   

For this simulation, the calculation started from a standard {\sc fastwind} model 
with a "$\beta$" velocity law and consisted in total of 345 NLTE and CMF 
radiative transfer iterations for computing 
the radiative acceleration, corresponding here to 35 
hydrodynamic iteration updates. The converged final model shows 
$\rm \log_{10} \Delta T_i = -4.3$, $\rm \log_{10} \Delta \varv_i = -2.5$, 
$\rm \log_{10} \Delta \varv_{\infty,i} = -3.7$, $\rm \log_{10} \Delta \dot{M}_i = -3.1$, 
and $\log_{10} f_{\rm err}^{\rm max} = -2.3$.  

\subsection{Influence of temperature structure} 
\label{tlucy} 

To test the influence of the simplified temperature structure 
eqn. \ref{Eq:Tsimple} upon the wind dynamics, this 
section computes a model with identical 
input parameters as for the early O-star in Table 1, but now using a  
full flux-conservation+electron thermal balance method (see above) 
to derive the run of the temperature. Indeed, also this model can 
be brought to convergence, but only after more than twice the number of NLTE 
iterations and a careful set-up of the initial "$\beta$-law" start-model. 
Nonetheless, this early O-star model converges to the same 
mass-loss rate $\dot{M} = 1.5 \times 10^{-6} \, \rm M_\odot/yr$ and 
terminal wind speed $\varv_\infty = 2500 \, \rm km/s$ as previously. 
Fig. \ref{Fig:t_comp} plots 
the temperature and velocity structures for the two converged models,  
demonstrating that although the temperatures as expected reveal 
some differences (e.g., a small reversal is now visible; see 
discussion in \citealt{Puls05} for more details about this feature), the resulting 
velocity fields are remarkably similar. 

And we can also quite readily understand this result by: i) noting that 
although $g_{\rm rad}$ in the optically thick, quasi-static layers 
is strongly dependent on the local gas temperature, in 
the thinner supersonic regions it becomes almost 
independent of it, and ii) inspecting the force balance at the sonic 
point and beyond, which reveals that in those regions 
the "Parker terms" $2 a^2/r -da^2/dr$ are less than a 
percent of the corresponding values of $g_{\rm rad}$ 
(i.e., for the supersonic part the e.o.m. is completely 
dominated by inertia and the counteracting radiative and 
gravitational accelerations). 

\subsection{Influence of wind clumping and X-rays}
\label{clumping} 

All models presented here solve the e.o.m. for a steady outflow. 
However, linear perturbation analysis shows the observer's frame 
line force is subject to a very strong instability on scales smaller than the Sobolev length  
 \citep{Owocki84}. Time-dependent models \citep{Owocki88, Feldmeier97, 
Sundqvist18} following the non-linear evolution of this line-deshadowing instability 
(LDI) show a characteristic two-component-like structure consisting of 
spatially small and dense clumps separated by large regions of 
very rarified material, accompanied by strong thermal shocks and 
a highly non-monotonic velocity field. But as shown, e.g., 
in Fig. 5 of \citet{Sundqvist18}, the averaged density and 
velocity in such models still exhibit a smooth and steady behavior. As 
such, for computations aiming at predictions of global quantities, such as 
mass-loss rates and terminal wind speeds, it may still be 
reasonable to solve the hydrodynamic equations in 
the steady limit. Nevertheless, the overdense clumps and the high-energy radiation 
resulting from the wind shocks may still affect the overall wind ionization 
balance \citep[e.g.,][]{Bouret05} and $g_{\rm rad}$, and so might induce feedback-effects  
also upon global quantities. 

Here we examine such feedback effects by computing models with 
identical input parameters as for the early O-star in Table 1, but now 
including parametrized forms of such wind clumping and X-ray emissions. X-rays are 
incorporated according to \citet{Carneiro16} and clumping according 
to \citet{Sundqvist18b}. Note though, that in this paper we restrict 
the analysis to clumps that are optically thin; potential \textit{dynamical} 
effects of porosity in physical and velocity space \citep{Muijres11, Sundqvist14} 
will be examined in a forthcoming paper. This assumption 
makes the basic treatment of wind clumping here very similar to 
that in the alternative steady models by \citet{Sander17} (except for the radial 
distribution of clumping factors). For the models displayed 
in Fig. \ref{Fig:clump}, we have assumed that clumping starts at a velocity 
$\varv = 0.1 \varv_\infty$ and increases linearly in velocity until 
$\varv = 0.2 \varv_\infty$, outside which it remains constant at a value 
given by a clumping factor $f_{\rm cl} = \langle \rho^2 \rangle/\langle \rho \rangle^2 
= \rho_{\rm cl}/\langle \rho \rangle =10$, for 
clump density $\rho_{\rm cl}$ and mean wind density 
$\langle \rho \rangle = \dot{M}/(4 \pi \varv r^2)$. While both 
observations \citep{Puls06, Najarro11} and theory 
\citep{Sundqvist13} indicate that clumping may be a more complicated 
function of radius, such a simple clumping-law should suffice for our 
test-purposes here. Note further 
that within the approximations used in this paper, $f_{\rm cl}$ is the inverse to the 
fractional wind volume $f_{\rm vol}$ occupied by clumps, i.e., here 
$f_{\rm cl} = f_{\rm vol}^{-1}$ (but see discussion in \citealt{Sundqvist18b}). 
X-ray emissions are assumed to start at $r \approx 1.5 R_\ast$ and 
follow the basic standard prescription outlined in \citet{Carneiro16}; the model 
presented in Fig. \ref{Fig:clump} has a resulting X-ray luminosity 
$L_{\rm x}/L_\ast \approx 0.8 \times 10^{-7}$. 

The lower panel of Fig. \ref{Fig:clump} compares $\Gamma$-factors 
from the simulation without clumping to i) one including only clumping but no 
X-rays and ii) one including both clumping and X-rays. As shown in the figure, 
for this particular model both these effects lead to a boost of the radiative 
acceleration in the outer parts of the wind. In turn then, this leads to higher wind speeds 
as illustrated by the upper panel of Fig. \ref{Fig:clump}. Due to the 
steep initial wind acceleration, the assumed start-velocity for clumping 
$\varv = 0.1 \varv_\infty$ corresponds to a rather low radii $r/R_\ast \approx 1.05$, 
implying clumping also in near-star wind regions. However, the 
mass-loss rates are only marginally different though, since $g_{\rm rad}$ 
around the critical point is barely affected by any clumping 
or X-ray feedback. The model including only clumping has a final 
$\dot{M} = 1.41 \times 10^{-6} \, \rm M_\odot/yr$ and $\varv_\infty = 3150 
\, \rm km/s$ and the model including both clumping and 
X-ray feedback results in 
$\dot{M} = 1.41 \times 10^{-6} \, \rm M_\odot/yr$ and $\varv_\infty = 3390 \, \rm km/s$. 

Note finally though, that future work should 
examine potential feedback-effects also from models that 
introduce clumping in even deeper and slower 
atmospheric layers, which may then affect also the 
predicted mass-loss rates.    

\subsection{Influence of turbulent speed} 
\label{vturb} 

The turbulent speed $\varv_{\rm turb}$ applied in our models 
(see $\S$\ref{rad_acc}) effectively broadens the line profiles and 
so also affects the computation of $g_{\rm rad}$. As also discussed
by, e.g., \citet{Poe90} and \citet{Lucy07a} this can have a significant impact upon 
the line acceleration in the critical sonic point region. 
More specifically, since a line profile extends over a few 
thermal+turbulent velocity widths, it means that, for our standard 
O-star value $\varv_{\rm turb} = 10$\,km/s, in the region leading up to the sonic point 
$a \approx 20 $\,km/s there is not enough velocity space available 
to completely Doppler shift line profiles out of their own absorption shadows. This 
then typically leads to a reduction of the line acceleration in these regions, as compared 
to Sobolev calculations where it is assumed that the line profiles are always 
fully de-shadowed so that the corresponding (fore-aft symmetric) line 
resonance zones do not extend into the stellar core. 
In turn, the reduced $g_{\rm rad}$ may then also reduce the 
predicted mass-loss rate as the overlying wind adjusts to the 
"choked" sub-sonic conditions.     

Fig. \ref{Fig:vturb} illustrates this effect, displaying results from models computed with identical 
parameters as the early O-star in Table 1, but with varying turbulent velocities 
$\varv_{\rm turb}$. The figure shows resulting mass-loss rates and terminal wind 
speeds, normalised to the standard model with $\varv_{\rm turb}=$\,10 km/s. 
The clearly visible trend in the figure confirms the above discussion; higher turbulent 
velocities mean lower predicted mass-loss rates. In addition, the figure shows 
that this then also means higher terminal wind speeds, essentially because 
the supersonic wind now needs to drive less mass off the stellar surface 
and so more easily can accelerate. Typical numbers, for this early O-star model,  
are a reduction in mass loss by approximately a factor of two when 
increasing $\varv_{\rm turb}$ from 10 to 18 km/s, accompanied  
by an increase in terminal wind speed by approximately 50 \%.  

\subsection{Late O-star model} 
\label{lateO} 

For the lower-mass model ($M \approx 27 M_\odot$), the upper two panels of Fig. 
\ref{Fig:O7} show the converged force balance analogous to Fig. \ref{Fig:force_bal}. 
At first glance, the lower-mass model displays similar characteristics as the higher-mass 
model, however closer inspection shows that the dip in $g_{\rm rad}$ in sub-sonic 
layers is now somewhat more prominent, and that once the Doppler-shifted  
line-opacity kicks in near the sonic point $g_{\rm rad}$ very quickly 
shoots up to values higher than 20 times the local gravity (cmp. Fig. \ref{Fig:force_bal} for 
the higher-mass model where this rise rather results in $\Gamma \approx 3$). 
The steep acceleration gives rise to a fast, low-density wind with a low final mass-loss 
rate $\dot{M} = 2.55 \times 10^{-8} \, \rm M_\odot/yr$ and high terminal wind speed 
$\varv_\infty = 5320 \, \rm km/s$, giving a scaled wind-momentum rate 
$\eta = \dot{M} \varv_\infty c/L_\ast = 0.05$.  

We can readily understand the 
character of the fast wind by considering the e.o.m. eqn. 
\ref{Eq:eom} in the supersonic, zero sound-speed limit:  

\begin{equation} 
	\varv \frac{d \varv}{dr} = \frac{G M_\ast (\Gamma-1)}{r^2}.
	\label{Eq:eom_ss}   
\end{equation} 

Noting then from Fig. \ref{Fig:O7} that  $\Gamma \approx Const. \approx 27$ 
after the initial steep rise close to the stellar surface $R_\ast$, 
eqn. \ref{Eq:eom_ss} can be analytically integrated to 

\begin{equation} 
	\varv(r) = \varv_{\rm esc}(R_\ast) \sqrt{\Gamma-1} \left( 1 - \frac{R_\ast}{r} \right)^{1/2},  
	\label{Eq:v_ana}  
\end{equation} 

yielding a terminal wind speed $\varv_\infty = \varv(r \rightarrow \infty) \approx 5300 \rm \,km/s$ for 
surface escape speed $\varv_{\rm esc}(R_\ast) = \sqrt{2 G M_\ast/R_\ast} = 1040 \, \rm km/s$. This 
agrees very well with the numerically computed value in Table 1. Moreover, eqn. 
\ref{Eq:v_ana} suggests that the supersonic wind here should be quite well described 
by a $\beta=1/2$ velocity law; this will be confirmed in $\S$\ref{beta}, where we compare 
our self-consistent simulations to models with such analytic $\beta$ fields.  

Further for this low-density wind model, only $\tilde{\tau}_F \approx 0.03$ at the sonic point and the stellar surface is located at a very low $\varv \approx 5.0 \times 10^{-3} \, \rm km/s$; indeed, Fig. \ref{Fig:O7} shows an overall sub-sonic region characterized by significantly lower velocities as compared to the early O-star 
model, reaching $\varv \approx 10^{-4} \, \rm km/s$ at the lower boundary 
$m_c^{\rm tot} = 80$. 

The iterative behavior of the hydrodynamic variables shows quite similar properties 
as for the early O-star model displayed in Fig. \ref{Fig:err_plot}; after an initial 
decrease in mass loss and increase in terminal wind speed, the simulation eventually 
finds its way towards convergence. We note though that numerically it is 
somewhat more difficult to make this model convergence, for example 
due to the sensitivity of the e.o.m. to 
the very steep acceleration around the sonic point. Fig. \ref{Fig:O7_ferr} shows 
$\dot{M}$ and the maximum error $f_{\rm err}^{\rm max}$ in the e.o.m. for each 
of the hydrodynamic iteration steps, demonstrating that although the simulation 
indeed reaches our convergence criteria, it does fluctuate a bit also for 
structures quite close to this. The final model shows 
$\rm \log_{10} \Delta T_i = -4.3$, $\rm \log_{10} \Delta \varv_i = -2.5$, 
$\rm \log_{10} \Delta \varv_{\infty,i} = -2.5$, $\rm \log_{10} \Delta \dot{M}_i = -2.4$, 
and $\log_{10} f_{\rm err}^{\rm max} = -2.2$.

\section{Discussion}
\label{discussion}

Having presented basic results and features above, this section now compares  
this paper's two core simulations to some other 
O-star wind models and results that appear in the literature. 

\subsection{Comparison to other steady-state models} 
\label{comparison} 

As mentioned in previous sections, the models by \citet{Sander17} 
are computed on similar assumptions as those adopted 
here (CMF radiative transfer, spherically symmetric steady-state hydrodynamics,  
grid-based $g_{\rm rad}$ depending explicitly only on radius). As such, 
we computed an additional model adopting the 
same stellar parameters, $T_{\rm eff} = 42\,000$\,K, $R_\ast/R_\odot = 15.9$, 
$\log_{10} L/L_\odot = 5.85$, $\varv_{\rm turb} = 15$\, km/s, as in \citet{Sander17}. Note, however, 
that, unlike \citet{Sander17}, the simulation here used a strict solar composition 
for the chemical abundances and also did not include clumping; nevertheless, 
the models are close enough to enable a quite fair and direct comparison. Indeed, 
our simulation converges to $\dot{M} = 1.58 \times 10^{-6} \, \rm M_\odot/yr$, 
$\varv_\infty = 2700 \, \rm km/s$, and $\eta = \dot{M} \varv_\infty c/L_\ast = 0.3$. 
For the mass-loss rate this is in perfect agreement with \citet{Sander17}, who 
also find $\dot{M} = 1.58 \times 10^{-6} \, \rm M_\odot/yr$. On the other hand, 
for the terminal wind speed (and thus the wind-momentum) they find slightly lower 
values, $\varv_\infty = 2000 \, \rm km/s$ and $\eta = 0.23$, than here. These 
differences may simply be an effect of the slight differences in input 
parameters and basic model assumptions (abundances, clumping, 
definition of stellar radius, see above and $\S$2). Overall the 
general agreement between the two independent calculations 
is encouraging, however future work should investigate in more detail the  
impact of, e.g., wind clumping upon the wind parameters 
predicted by these CMF calculations (see also $\S$\ref{clumping}).   

Also \citet{Krticka10, Krticka17} use CMF transfer to derive the radiative acceleration 
in their simulations. However, these models scale the CMF line force to a 
corresponding Sobolev-based one, which effectively means the critical point in their e.o.m. 
is shifted upstream from the sonic point to the supersonic point 
first identified by CAK \citep{Krticka17}. As such, their 
models may potentially have quite different convergence behavior 
than the simulations presented here. Nonetheless, 
inserting the stellar luminosities of Table 1 into the mass-loss scaling 
relation eqn. 11 in \citet{Krticka17} yields $\log_{10} \dot{M} = -5.9$ and  
$\log_{10} \dot{M} = -7.2$  for the models with higher and lower 
luminosities, respectively. For the former this agrees well with the rate 
derived here, whereas for the latter we predict a significantly 
lower value (see Table 1). Recalling again that there are some important differences in 
modeling techniques between \citet{Krticka10,Krticka17} and this paper, further 
studies are required to trace the exact origin of this. 

The most widely used mass-loss rates for applications like stellar 
evolution are those compiled by \citet{Vink00,Vink01}. These are 
based on global energy considerations, using a prescribed 
$\beta$ velocity law and a Monte-Carlo line force based on 
approximate NLTE computations and the Sobolev approximation. 
An advantage with this approach is that since the vast majority of 
the radiative energy is transferred to the wind in the supersonic 
regions (where the Sobolev approximation is valid), the method does not 
depend sensitively on the transonic flow properties (nor on the 
turbulent speed there, see \S \ref{vturb} and below). Moreover, since 
the velocity field is parametrized $\varv_\infty$ can 
readily be adjusted (e.g., to an observationally inferred value) 
when deriving the corresponding mass-loss rate.
Using $Z/Z_\odot=1$ and 
the recommended $\varv_\infty = 2.6 \varv_{\rm esc}^{\rm eff}(R_\ast)$ 
in the Vink et al. (2000) mass-loss recipe, 
we find $\log_{10} \dot{M} = -5.3$ and $\log_{10} \dot{M} = -6.6$ for our  
early and late O-stars, respectively. This is a factor of 
$\sim 3$ and $\sim 9$ higher than the rates derived in this paper.
By accounting for our models' higher $\varv_\infty$, the agreement 
can be somewhat improved upon. Moreover, we note that the solar 
base metallicity is somewhat lower here than in Vink et al., simply because they 
used solar abundances by \citet{Anders89} ($Z_\odot = 0.019$) 
whereas this paper assumes those by \citet{Asplund09} ($Z_\odot = 0.013$). 
However, a simple scaling using $Z = 0.013/0.019$ in the \citet{Vink01} 
recipe overestimates the effect (since the base-abundance of important 
driving ions like iron has not changed much). To 
quantify (see also \citealt{Krticka07}), we ran an additional model with the same 
parameters as the early O-star in Table 1, but now assuming 
solar abundances according to \citet{Anders89}. This simulation converged 
to a mass-loss rate $\sim 20$\,\% higher than our base model (as 
compared to $\sim 40$\,\% if applying the simple metallicity scaling 
above). As such, also with these metallicity modifications a significant 
discrepancy remains between our rates and the Vink et al. scalings
(a factor $\sim 2.5$ in the early O-star case). Most likely these differences 
arise due to a combination of the use 
here of CMF transfer instead of the Sobolev approximation 
in the critical near-star regions (see $\S$\ref{vturb} and also, e.g., \citealt{Owocki99}) and the 
different NLTE solution techniques used here and in Vink et al. As 
mentioned above, the global Sobolev Monte-Carlo method 
utilized by Vink et al. is not directly very sensitive to the conditions 
around the critical sonic point. However, since these models 
are parameterized to follow a $\beta$ velocity law, the method will 
nevertheless indirectly neglect the effects the local reduction of the near-star 
CMF radiation force have upon also the supersonic velocity field 
and the global energy balance. Regarding the 
non-Sobolev models by \citet{Lucy07a, Lucy10}, these 
parametrize the Monte-Carlo line force as purely a function of velocity, 
and further solve the e.o.m. only in the plane-parallel limit 
up to a velocity $\approx 4 a$ (i.e., focusing on the wind initiation only).
In any case, also \citet{Lucy10} reports on significantly lower mass-loss 
rates than those obtained by the corresponding Vink et al. recipe, 
in general agreement with the above. 

Finally, regarding the dependence of the mass-loss rate 
on the microturbulent speed (\S \ref{vturb}), we note that both 
\citet{Lucy07a} and \citet{Krticka10} identify the same 
qualitative behavior as here, namely somewhat 
lower rates for higher $\varv_{\rm turb}$.        

\subsection{Comparison to "$\beta$-law" models} 
\label{beta} 

As mentioned above, in the standard version of {\sc fastwind}
(as well as in other codes used for quantitative spectroscopy of hot, massive stars),  
an analytic "$\beta$-law" is typically smoothly combined with the quasi-static photosphere 
in order to approximate the outflowing wind velocity field. Fig. \ref{Fig:beta} compares 
models computed with such $\beta$-laws to the self-consistent velocity 
structures above. To allow for simple visual inspections, it is here useful to 
re-cast the radial coordinate in a dimensionless parameter $X = 1-r_c/r$, 
such that $\beta$ in the high-velocity parts simply becomes the slope on 
a log-log plot showing velocity vs. radius, i.e., 
$\log_{10} \varv/\varv_\infty =  \beta \log_{10} X$. 

The figure shows fits to the self-consistent velocity structures using a "single" $\beta$-law 
of the form:  

\begin{equation} 
		 \varv(r) = \varv_\infty u^\beta \ \ \ \ \ \ \ \ \ \    u = 1-\frac{r_{\rm tr} \ b}{r}
		 \label{Eq:beta1} 
\end{equation}  

with transition radius $r_{\rm tr}$ and $b = 1 -(\frac{\varv_{\rm tr}}{\varv_\infty})^{1/\beta}$ derived 
from a transition velocity $\varv_{\rm tr}$ (see also $\S$\ref{numerical}, 
first paragraph). For comparison, we also display fits using "double" $\beta$-laws of the form: 

\begin{equation} 
		 \varv(r) = \varv_\infty \left( (1-u) \ u^\beta + u \ u_2^{\beta_2} \right) 
                  \label{Eq:beta2} 
\end{equation}     

for  $u_2 = 1-r_{\rm tr} \ b_2/r$ and 
$b_2 = 1 -(\frac{\varv_{\rm tr}}{\varv_\infty})^{1/\beta_2}$. 
Such "double" $\beta$-laws are sometimes also used in spectroscopic wind 
studies, for example in attempts to better capture the different 
acceleration regions in WR-stars. 

Using eqn. \ref{Eq:beta1} (eqn. \ref{Eq:beta2}), 
Levenberg-Marquardt best-fits for the parts 
with $\varv > \varv_{\rm tr}$ were created by varying $\beta$ (and $\beta_2$) 
for a range of assumed $\varv_{\rm tr}$. For the early O-star model, the best-fit single-law model shows 
$\beta = 0.68$ and $\varv_{\rm tr}/a(T_{\rm eff}) = 0.54$. This velocity field is compared to the 
self-consistent model in the lower panel of Fig. \ref{Fig:beta} (dashed line), showing 
good agreement in particular for the highly supersonic parts. A transonic 
point $\varv \approx 0.5 a(T_{\rm eff})$ is significantly higher than 
the $\varv \approx 0.1 a(T_{\rm eff})$ typically used in {\sc fastwind}, and needed here 
to prevent the $\beta$ velocity field from shooting up too early.   
This indicates that the wind acceleration sets in at relatively high subsonic velocities in 
the self-consistent models. Indeed, for the late O-star simulation the best-fit 
single-law has $\beta=0.57$ and an even higher $\varv_{\rm tr}/a(T_{\rm eff}) = 0.8$. 
The upper panel of Fig. \ref{Fig:beta} compares this velocity field to the 
hydrodynamic one, confirming the earlier discussion in 
$\S$\ref{lateO} that the near constancy of $\Gamma$ in the supersonic 
parts here implies a steeper $\beta \approx 0.5$. 

It is further interesting to note that the double $\beta$-laws (dashed-dotted lines in 
Fig. \ref{Fig:beta}) actually do not provide significantly better fits than the simpler single-laws discussed above. 
Namely, while for this more complex parametrization $\beta$ remains well-constrained,  
and basically unaffected from the above, our fits also signal that while for the 
early O-star some constraints can be placed on the $\beta_2$ parameter, for 
the late O-star $\beta_2$ is basically unconstrained (no visual improvements are seen to the fits 
either, see the dashed and dashed-dotted curves in Fig. \ref{Fig:beta}). This 
suggests that this double $\beta$-law is not an optimal form 
for characterization of the present O-star hydrodynamic velocity structures. Rather 
(as also can be visually seen in Fig. \ref{Fig:beta}), the main problem with 
the single $\beta$-law regards capturing the strong velocity rise and curvature 
in near-star regions. In this respect, some first tests suggest that a promising way may be to
in the low-velocity wind regions also account for a quasi-static exponential rise 
$\varv \sim e^{\Delta r/h}$, with some effective scale height $h$, in the 
above parameterizations. This idea will be further explored in the 
follow-up paper, where a larger grid of hydrodynamic 
models to compare with will be presented. 

Overall, these first comparisons thus suggest that for early O-stars a model 
with a photosphere to wind transition point $\varv \approx 0.5 a(T_{\rm eff})$ 
accompanied by a $\beta \approx 0.7$ law may be a quite good initial choice for 
quantitative spectroscopic studies not aiming for theoretical predictions of the 
global wind variables, but rather at empirical derivations of them. On the other 
hand, for late O-stars with low-density winds a steeper 
$\beta \approx 0.5-0.6$ seems to be a 
better choice, at least for the highly supersonic parts.  

\subsection{Comparison to CAK line force} 

In CAK-theory \citep{Castor75}, the radiation line force 
in a spherically symmetric wind is:  

\begin{equation} 
	g_{\rm cak}  = \frac{g_{\rm e} \bar{Q}}{ (1-\alpha) \, (\bar{Q} \kappa_e \rho c/d\varv/dr)^\alpha } f_d
	\label{Eq:gcak} 
\end{equation} 

where $\alpha$ is CAK's power-law index, which 
physically describes the ratio of the optically thick line force contribution to 
the total one \citep{Puls00}, 
and $f_d$ is the finite-disc correction factor \citep{Pauldrach86, Friend86}:  

\begin{equation} 
	f_d = \frac{ (1+\sigma)^{1+\alpha} - (1+\sigma \mu_\ast^2)^{1+\alpha} } 
	{ (1+\alpha) \sigma (1+\sigma)^\alpha (1- \mu_\ast^2) } 
\end{equation} 		 

for $\sigma = d \ln \varv / d \ln r-1$ and $\mu_\ast^2 = 1 - R_\ast^2/r^2$. 
In eqn. \ref{Eq:gcak} $\bar{Q}$ is the line strength normalization 
due to \citet{Gayley95}, representing the ratio between the line force and the 
electron scattering force $g_{\rm e}$ in the case all lines would be 
optically thin (here, $\alpha=0$). 
For a given $\alpha$, we may use these CAK expressions to compute what
"effective" $Q_{\rm eff}$ the CMF line force $g_{\rm cmf}$ ($g_{\rm rad}$ 
reduced by the continuum acceleration) corresponds to:  

\begin{equation} 
	Q_{\rm eff} = \left( \frac{g_{\rm cmf} (1-\alpha) } { g_{\rm e} f_d } 
	\left( \frac{ \rho \kappa_{\rm e} c}{d\varv/dr} \right)^\alpha \right)^{1/(1-\alpha)}. 
	\label{Eq:qeff} 
\end{equation}  

Fig. \ref{Fig:qeff} shows this $Q_{\rm eff}$ as function of velocity and dimensionless radius $X$
(see above), for the converged early O-star model in Table 1 and 
using a prototypical $\alpha = 2/3$ \citep{Puls00}. Because CAK-theory 
uses the Sobolev approximation to derive the line force, only the 
supersonic parts are displayed. The figure shows how $Q_{\rm eff}$ in the CMF model 
increases from about 1000 at the base to about 2000 at  $\varv \approx 100$\,km/s, 
after which it again starts to decline toward larger radii. The decline in 
$Q_{\rm eff}$ when approaching the stellar surface here reflects the very steep acceleration 
around the sonic point seen in our CMF models, which implies very high velocity 
gradients $d\varv/dr$ and so lower values of  $Q_{\rm eff}$ for a 
fixed $\alpha$ (see eqn. \ref{Eq:qeff}).  

It is interesting to note that these 
"effective" values indeed agree quite well both with the original recommended 
$\bar{Q} \approx 2000$ by \citet{Gayley95} and with the line statistics 
computations by \citet{Puls00}. In addition, the comparisons 
provide some support to radiation-hydrodynamical  
simulations that, for early O-stars in the Galaxy, typically use a 
radially fixed $\bar{Q} = 2000$ when modeling the time-dependent 
dynamical evolution of the wind \citep[e.g.,][]{Sundqvist14, Sundqvist18}. 

We caution, however, that a similar analysis for 
the lower-mass, late-type O-star would result in significantly lower $Q_{\rm eff}$
values (reaching maximum values of only a few hundred if again 
assuming $\alpha=2/3$), due primarily to the very low mass-loss rate and steep initial 
acceleration found for this star (see previous sections and eqn. \ref{Eq:qeff}).
In this respect, however, it is important to point 
out that the CMF line force here is computed quite differently than 
$g_{\rm CAK}$ in CAK-theory (not using the Sobolev approximation nor any assumptions 
about underlying line-distribution functions). In particular, $g_{\rm cmf}$ here 
has no explicit dependence on velocity gradient or velocity, in contrast to 
$g_{\rm CAK}$ (eqn. \ref{Eq:gcak}). As discussed below, this can 
have important consequences not only for direct comparisons 
of models, but also for the very nature of the steady wind solution.    

\subsection{Stability, uniqueness, and wind topology}
\label{topology} 
 
The two base models presented above show a quite 
stable iteration behavior toward convergence. However, in our 
calculations we have found that this is not always the case. For 
some models, the iteration rather first proceeds as expected,  
toward lower and lower $f_{\rm err}^{\rm max}$ and toward a more 
and more constant $\dot{M}$, but then never reaches the 
(admittedly rather strict) convergence criteria discussed above. Instead, 
the model may experience an almost semi-cyclic behavior, 
wherein low values of $f_{\rm err}^{\rm max}$ can be observed
during semi-regular iteration intervals, but where these minima never 
reach our selected criterion for convergence. To this end, it is neither 
clear what exactly causes this behavior, nor if it is a 
numerical or physical effect (see further below). 

For such simulations, for example the ones 
with the lowermost turbulent velocities displayed in Fig. 
\ref{Fig:vturb}, we then instead use the iteration with the 
lowest $f_{\rm err}^{\rm max}$ when selecting our final model. 
Note that while these simulations thus are not formally 
converged, the range in the most important targeted 
parameter $\dot{M}$ with acceptably low 
$f_{\rm err}^{\rm max}$ is typically quite modest.  
Specifically for the case mentioned above, the lowest
$f_{\rm err}^{\rm max}$ was on the order of a few 
percent, making it above our convergence criterion 
here but, e.g., below that used by \cite{Sander17} (5 \%); 
the corresponding mass-loss rates having 
$f_{\rm err}^{\rm max} < 10 \, \%$ lied in the 
range $\approx (1.5-2.0) \times 10^{-6} \rm \, M_\odot/yr$.  

Indeed, since the steady-state e.o.m. eqn. \ref{Eq:eom} is 
non-linear, it is not a priori certain that a unique and stable solution 
can be found. For time-dependent wind models, computed 
using distribution functions and an observer's frame line 
force, this has been analyzed by \citet{Poe90} 
and \citet{Sundqvist15}. For their parametrized line forces, 
these authors carried out topology analyses of the solution 
behavior at the critical sonic point\footnote{Since the velocity gradient 
does not appear explicitly in such observer's frame calculations 
of $g_{\rm rad}$, the sonic point remains the only critical 
point in the steady limit.}. It turns out that, depending for 
example on the behavior of the source 
function at the sonic point \citep{Sundqvist15}, the 
wind topology can transition from a stable X-type 
with a unique solution to an unstable nodal type 
exhibiting degenerate solutions. When such 
a wind model lies on the shallow part of this nodal topology branch, 
\citet{Poe90} showed that an iteration scheme like that applied 
here on the time-independent e.o.m. will not converge. 

In this respect, it is important to point out that since the models here 
compute $g_{\rm rad}$ as an explicit function only of radius (thus 
not including any explicit velocity dependence), we are effectively forcing 
the solution to lie on the stable X-type branch, irrespectively of 
what the "true" underlying solution actually is. Indeed, such
forcing toward the X-type branch is effectively done also in the alternative models by 
\citet{Sander17} and \citet{Lucy07a}, where the first use a line 
acceleration depending only on radius (like here) and the latter one depending only on 
velocity (thus neglecting the radial dependence). 

In any case, further topology analysis of numerical CMF models 
such as those presented here would require a completely 
new parametrization of the line force, in order to 
examine properly the solution behavior for different stellar 
ranges. This lies beyond the scope of the present paper, but should  
be a focus for follow-up work targeting detailed 
comparisons of our new steady-state models with 
time-dependent hydrodynamic calculations.     

\subsection{The "weak-wind" problem} 

The low mass-loss rate derived for the late O-star in $\S$\ref{lateO} is generally consistent with the 
empirical finding that observationally inferred mass-loss rates of late O-dwarfs typically are 
much lower than previous theoretical models suggested 
(the so-called "weak-wind" problem, see, e.g., \citealt{Puls08}; 
\citealt{Marcolino09}). Indeed, for our late-O model star we obtain 
a $\sim$9 times lower rate than that predicted by the theoretical mass-loss 
recipe by \cite{Vink00}, as already discussed in $\S$\ref{comparison}. 

On the other hand, the accompanying very high terminal 
wind speeds we derive have not been observed. This suggests that some additional 
physics may act to reduce the radiation force in the outer parts of these low-density 
winds from Galactic O-dwarfs (or at least make them invisible to UV observations). 
In this respect, one speculation is that LDI-generated shocks (see $\S$\ref{clumping}) 
might not be able to efficiently cool in such low-density environments, so that also 
the bulk wind is left much hotter than expected. Such a shock-heated bulk wind 
seems to be suggested by the X-ray observations of low-density winds by 
\citet{Huenemoerder12} and \citet{Doyle17}, who also suggested that the 
"weak-wind" problem might simply be related to insufficient cooling (see also 
\citealt{Lucy12}). Indeed, from soft X-ray {\sc rosat} observations 
already \citet{Drew94} suggested that large portions of hot, uncooled gas 
might exist in the outer regions of low-density winds. If 
true, such extended hot regions could then also imply a significantly 
reduced radiative driving in these wind parts. 

Note that this scenario is quite different from that examined in $\S$\ref{clumping} for 
the early O-star wind with a higher density.  
Namely, although the included X-rays there may have a significant impact on 
the ionization balance of some specific elements, the temperature structure 
of the bulk wind is basically unaffected. The possibility of a hot bulk wind should be further 
examined in future work, preferably by performing LDI simulations tailored to low-density 
winds and including a proper treatment of the shock-heated gas and associated
cooling processes. 

\section{Summary and future work} 
\label{summary} 

We have developed new radiation-driven wind models for hot, 
massive stars, suitable for basic predictions of global  
parameters like mass-loss and wind-momentum rates. The 
simulations are based on an iterative grid-solution to the 
spherically symmetric, steady-state e.o.m., using full NLTE CMF 
radiative transfer solutions for calculation of the radiative 
acceleration responsible for the wind driving. 
 
In this first paper, we present models representing two prototypical 
O-stars in the Galaxy, one "early" with a higher mass 
$M_\ast/M_\odot \approx 59$ and one "late" with a lower 
mass $M_\ast/M_\odot \approx 27$. Table 1 summarizes the results, 
with predicted mass-loss rates $\dot{M} = 1.51 \times 10^{-6} \, \rm M_\odot/yr$ 
and $\dot{M} = 2.55 \times 10^{-8} \, \rm M_\odot/yr$ for the two models, 
respectively. In $\S$4, we further discussed in some detail the influence on 
the model predictions from additional parameters like wind clumping, 
turbulent speeds, and choice of temperature structure calculation.   

Overall, our results seem to agree 
rather well with other steady-state wind models that 
do not rely on the Sobolev approximation for computation of the radiative 
line acceleration \citep{Lucy10, Krticka17, Sander17}. A key result
is that the mass-loss rates we predict are significantly lower than 
those \citep{Vink00, Vink01} normally included in models of 
massive-star evolution. Two likely reasons for the lower values are 
the use of CMF transfer vs. the Sobolev approximation and also the somewhat 
different nature of the NLTE solution techniques used here and by Vink et al. 
Our results thus add to some previous notions \citep[e.g.,][]{Sundqvist11, Najarro11, 
Surlan13, Cohen14, Krticka17, Zsolt17} that O-star mass-loss rates may be 
overestimated in current massive-star evolution models and 
that (considering the big impact of mass loss on the lives of massive 
stars) new rates might be needed. 

In this planned 
paper series, we intend to work toward providing 
such updated radiation-driven mass loss for direct use in 
stellar evolution calculations  
(Bj{\"o}rklund et al., in prep.). Other follow-ups include 
investigations of the mass-loss vs. metallicity dependence in 
our models, comparison to time-dependent simulations, and 
extension of the current O-star models to B-stars (in particular 
supergiants across the 
bistability jump) and potentially also Wolf-Rayet stars.  
 
\begin{acknowledgements}

We thank the referee for useful comments on the manuscript. 
We also thank Stan Owocki for so many fruitful discussions 
on line-driving. JOS and RB also thank the KU 
Leuven {\sc equation}-group for their valuable 
inputs as well as their weekly supply of fika, and 
also acknowledge the Belgian Research Foundation 
Flanders (FWO) Odysseus program under grant 
number G0H9218N. JOS further
acknowledges previous support from the European 
Union Horizon 2020 research and innovation program under the 
Marie-Sklodowska-Curie grant agreement No 656725. 
FN acknowledges 
financial support through Spanish grants ESP2015-65597-C4-1-R 
and ESP2017-86582 C4-1-R (MINECO/FEDER).
 
\end{acknowledgements}

\bibliographystyle{aa}
\bibliography{sundqvist_mdot_cmf}

\end{document}